\documentclass[twocolumn]{aastex701}
\graphicspath{{./}{figures/}}

\usepackage{xcolor}
\usepackage{soul}

\usepackage{ulem}
\usepackage{amsmath}

\definecolor{forestgreen(web)}{rgb}{0.13, 0.55, 0.13}

\usepackage{subcaption}

\usepackage{tikz}
\usetikzlibrary{3d}
\usetikzlibrary{arrows.meta}

\begin{document}

\title{Direct Numerical Comparison of Data Driving Strategies for Solar Flux Emergence in Magnetohydrodynamics: $\mathbf{B}$-driving vs $\mathbf{E}$-driving 
	\footnote{Direct comparisons of magnetic field and electric field data-driving strategies in the Athena++ MHD solver}
}


\author[orcid=0000-0001-8217-8029,sname='Tapinou']{Kyriakos Christos Tapinou}
\affiliation{Sydney Institute for Astronomy, University of Sydney}
\email[show]{kyriakos.tapinou@sydney.edu.au}  

\author[orcid=0000-0000-0000-0001,sname='Raboonik']{Axel Raboonik}
\affiliation{School of  Science, University of Newcastle, Callaghan, Australia}
\email[hide]{axel.raboonik@newcastle.edu.au}

\author[orcid=0000-0000-0000-0001,sname='Pontin']{David I. Pontin}
\affiliation{School of  Science, University of Newcastle, Callaghan, Australia}
\email[hide]{David.Pontin@newcastle.edu.au}

\author[orcid=0000-0001-5100-2354,sname='Wheatland']{Michael S. Wheatland}
\affiliation{Sydney Institute for Astronomy, University of Sydney}
\email[hide]{michael.wheatland@sydney.edu.au}

\author[orcid=0000-0003-2110-9753,sname='Cheung']{Mark C.M. Cheung}
\affiliation{CSIRO Space and Astronomy, P.O. Box 76, Epping, NSW 1710, Australia}
\email[hide]{mark.Cheung@csiro.au}

\author[orcid=0000-0000-0000-0001,sname='Schunker']{Hannah Schunker}
\affiliation{School of  Science, University of Newcastle, Callaghan, Australia}
\email[hide]{hannah.schunker@newcastle.edu.au}


\begin{abstract}




\emph{Data driving} is an important technique for modelling realistic magnetic field and plasma evolution in regions of the solar atmosphere where direct observations are either absent or insufficient. 
This technique uses observational \emph{data} to \emph{drive} the evolution of the simulation, while the modelled solar atmosphere evolves self-consistently. 
A common approach for data-driven simulations is $\mathbf{B}$-driving, which uses the observed magnetic field at the surface of the Sun as a boundary condition. 
However, only prescribing $\mathbf{B}$ at the boundary does not generally determine its numerical evolution when the update depends on the reconstructed interface states or fluxes. 
To address this, $\mathbf{E}$-driving provides an alternative approach by instead driving the simulation using the electric field on the boundary, which directly controls the magnetic field update. 
We systematically quantify the performance of $\mathbf{B}$- and $\mathbf{E}$-driving in a controlled set of numerical experiments.
We test both methods in the Athena++ solver using (a) an analytical flux-emergence model \citep{fan2003emergence} and (b) a ``ground-truth'' reference simulation \citep{toriumi2017numerical,toriumi2020comparative} of active region emergence. 
For $\mathbf{E}$-driving, the electric field is incorporated into Athena++'s constrained-transport (CT) update, preserving the discrete solenoidal constraint. 
We assess the reproduction of (i) boundary magnetic flux, (ii) relative magnetic helicity, (iii) total and free magnetic energy, and (iv)  magnetic morphology. 
Across these diagnostics, $\mathbf{E}$-driving gives better overall reproduction of the reference evolution than $\mathbf{B}$-driving.
For the configurations considered in an Eulerian CT scheme, $\mathbf{E}$-driving is more accurate than the commonly used ghost-cell $\mathbf{B}$-driving strategies and simpler than constructing an update-consistent $\mathbf{B}$-driving boundary condition.

\end{abstract}


%
%
%


\section{Introduction}


Data-driven numerical simulations are a powerful approach for realistic modelling of the solar atmosphere \citep{cheung2012method,hayashi2018mhd,cheung2019comprehensive, jiang2022datadriven, jiang2023data,chen2023data}. 
They can be thought of as time-dependent boundary value problems, whereby an assumed initial atmospheric state is evolved using (observed or synthetic) dynamic data at the photosphere \citep{Wu_2006}. 
In recent years, various studies have used this approach to model solar eruptive events such as flares and coronal mass ejections (CMEs) via data-driven flux emergence simulations  \citep[e.g.,][]{hayashi2018mhd,guo2019solar,guo2023thermodynamic,jiang2023data,guo2024magnetic,fan2024data,chen2026data,guo2026data}. 
Solar eruptions play a central role in shaping space weather, making their prediction using time-efficient observation-informed numerical modelling an imperative \citep{Jiang2016,Vourlidas2019}.
%


As a dynamic technique, data driving incorporates evolving boundary conditions derived from time series of high spatial resolution maps (such as vector magnetograms and velocity maps). This can be contrasted with static techniques such as extrapolation and data-constrained simulations, which typically rely on a single snapshot of (boundary) data.
Thus, data-driven simulations offer realistic physics-grounded coupling between the lower and upper solar atmosphere. 
In particular, such simulations have been successful in reproducing important observed magnetic signatures including the buildup of non-potential magnetic energy and helicity in the corona \citep{cheung2014flux}, as well as shearing driven by differential rotation in coronal arcades \citep{bhowmik2021two}.
Thus, the goal of data-driving is the reconstruction and evolution of the magnetic field and plasma in the corona above realistic active regions, which is essential for modelling solar eruptions, coronal heating, and the solar wind.

One way to generate magnetic reconnection is to implement flux emergence into an overlying existing magnetic field and a number of reduced data-driven models have been employed to simulate active region eruptions through flux emergence, including magneto-frictional \citep{cheung2012method}, zero-$\beta$ magneto\-hydro\-dynamic (MHD) \citep{guo2019solar}, and isothermal MHD \citep{jiang2023data} models. While such simplified models are more tractable and require less computational resources, they 
omit some of the physics necessary to capture the full plasma evolution. 
Full-MHD models are able to reproduce key eruption features such as flare and CME onset times, as well as thermal emission profiles \citep{hayashi2018mhd,jiang2023data,chen2023data,guo2023thermodynamic,fan2024data}. However, it remains a challenge to determine the best way to incorporate realistic active regions into these models~\citep[see review by][]{guo2024magnetic}.

The most straightforward data-driving method is magnetic field driving (hereafter $\mathbf{B}$-driving), where the boundary conditions (BCs, formal explanation is given in Section~\ref{sec:frame_data_driving}) on the magnetic field (or at least its normal component) are evolved from observed vector magnetograms. 
Near-surface convective flows are thought to play an important role in the flux emergence process \citep[e.g.][]{Birch+2019,Schunker+2024}, typically advecting the magnetic field on timescales of weeks to months (longer in some cases) \citep{yeates2023surface}. 
The corresponding plasma flows at the bottom  boundary can be specified either through ad hoc BCs (e.g.\ zero-gradient \citep{guo2019solar,guo2024magnetic,Guo2024FluxRope} or zero-mass flux \citep{chen2023data,chen2026data}) or they can also be directly specified from the observations 
\citep[e.g.][]{Schuck_2008}. 
This completes the necessary BCs to realistically drive flux transport and shear injection \citep{Fisher_2010, hayashi2018mhd}.
However, the direct imposition of external non-self-consistent BCs requires careful consideration of the numerical implementation to solve and satisfy the physical conditions. 

Imposing $\mathbf{B}$-driving requires assumptions about the vertical gradient of the field across the bottom boundary. 
Another data driving approach is to prescribe the electric field at the photospheric boundary (hereafter $\mathbf{E}$-driving), which does not need to rely on these assumptions. 
This approach requires an MHD solver in which the electric field, or its discrete electromotive force (EMF) representation, can be prescribed directly in the induction equation update.
Moreover, calculating the electric field from observations is itself a major source of uncertainty. 
This has led to the development of a suite of electric field inversion techniques as a viable  alternative \citep{fisher2010estimating, cheung2012method, kazachenko2014comprehensive, cheung2015homologous,weinzierl2016new,pomoell2019time, price2019time}. 
In the simplest scenario, assuming complete knowledge of the photospheric magnetic field and flow velocities (as well as currents and resistivity in the case of non-ideal MHD), one can use Ohm's law to calculate $\mathbf{E}$. 
This is straightforward when using synthetic data-driving BCs from simulations, however when using observations the data completeness assumption is usually not met, and the missing quantities must be inferred.
One example is the Poloidal–Toroidal Decomposition (PTD) scheme, which breaks down the electric field into inductive (divergence-free) and non-inductive (curl-free) components. We will quantify the effectiveness of this approach in this paper.

The choice of which data-driving strategy to adopt should be motivated by the reliability of the simulation in order to produce results that are faithful to both the input data and the MHD model. 
Such a strategy relies on: the numerical solver that updates the magnetic field in the simulation given a new BC; the availability and quality of the observational data to be used as BCs; and the physics priorities. 
In this paper we quantitatively evaluate the success of $\mathbf{B}$-driving and $\mathbf{E}$-driving algorithms in modelling flux emergence.

Since observations cannot provide a complete set of boundary conditions
\citep{Leake_2017,toriumi2020comparative}, we first use the complete surface magnetic field and flow quantities from an externally computed simulation of magnetic flux as the BC for our simulation.
Notably, \cite{Leake_2017} showed that if all of the MHD state variables are known 
over multiple vertical grid-points representing the photosphere (or bottom boundary),
then in principle the evolution in the volume above that surface can be reproduced. 
To quantify the success of data driving when using observations, in this paper we consider only a single plane of a ``ground truth'' simulation.
We select two numerical models to use as the ground-truth simulations:
\citet{fan2003emergence} simulated the steady emergence of an analytical toroidal flux-tube, while \citet{toriumi2017numerical} simulated the buoyant rise of a flux-tube embedded in the convection zone.
We implement the maps of the vector magnetic field and velocity at the surface of the Sun from these simulations as the BCs to drive our simulations, using the Athena++ simulation tool \citep{stone2008athena,stone2020athena++}, which is capable of implementing both $\mathbf{B}$- and $\mathbf{E}$-driven techniques. 
We formally quantify and validate the output of our simulations with the ground-truth simulations.
As a further validation, we also include a simple comparison between our results obtained by Athena++ and a $\mathbf{B}$-driving implemented LareXd \citep{ARBER2001151} using the same data to drive the simulations
\section{Data-driving terminology for solar physics}\label{sec:driving_vs_assimilation}

Data-driving in the solar physics literature does not use the standard terminology that has been developed in other fields. One exception is perhaps the ADAPT surface flux transport model~\citep{Hickmann:2015}, which is limited to modelling the evolution of the radial flux density $B_r$ at the photosphere.

Adopting standard terminology will allow the solar physics community to exploit the rich data-driving literature in other fields. 
To this end, we summarise some of the unifying notation for data-driving introduced by \citet{ide1997unified} in terrestrial weather forecasting. 

In the context of solar MHD simulations, the evolved MHD state is referred to as the \emph{realised state}, which can apply to the state on the evolved boundary and/or the state within the simulation domain. 
Notation (from \citet{ide1997unified}) used in this paper  are:
$\boldsymbol{x}(t_i)$ is the model's state vector at some discrete time $i$; $\boldsymbol{M}_i$ is the model's dynamics operator; $\boldsymbol{y}^{o}_i$ is the observation at $t_i$; $\boldsymbol{H}_i$ is the observation operator; $\boldsymbol{d}_i$ is the observational residual (also known as the innovation vector); and $\boldsymbol{R}_i$ is the covariance operator.
Superscripts $(\cdot)^\mathrm{f}$, 
$(\cdot)^\mathrm{a}$, and $(\cdot)^\mathrm{o}$ indicate quantities
associated with a forecast, 
analysis, and observation, respectively. The three main families of techniques commonly used in MHD simulations can then be classified as follows:
\paragraph{Data-agnostic schemes} In the absence of observational data, a discrete model for the evolution of a generic deterministic system is given by
\begin{eqnarray}\label{eq:ide_data_driving}
    \boldsymbol{x}^\mathrm{f}\left(t_{i+1}\right) &=& \boldsymbol{M}_i\left[\boldsymbol{x}^\mathrm{f}\left(t_i\right)\right]. 
\end{eqnarray}
Here, the realised (or forecast) state vector at $t_{i+1}$ is a function of its own previous realised state at $t_{i}$ alone.

\paragraph{Data-Driven (DD) schemes} In contrast, a data-driven discrete model is required to update the state vector not only through $\boldsymbol{x}^\mathrm{f}\left(t_{i}\right)$, but also in response to raw observational data $\boldsymbol{y}_i^\mathrm{o}$.
In solar MHD modelling, $\boldsymbol{y}_i^\mathrm{o}$ may represent, say, time-dependent photospheric magnetograms. The data-driven forecast state vector can then be generally written using a composite dynamical operator that is a function of both the previous time-step's realised state as well as observation, namely
\begin{eqnarray}\label{eq:data-driving}
    \boldsymbol{x}^\mathrm{f}\left(t_{i+1}\right) = \boldsymbol{M}_i\left[\boldsymbol{x}^\mathrm{f}\left(t_i\right),\boldsymbol{y}^\mathrm{o}_i\right]. 
\end{eqnarray}
The objective of data driving is to produce a realised state such that the model-forecast observation $\boldsymbol{y}_i=\boldsymbol{H}_i[\boldsymbol{x}^\mathrm{f}(t_i)]$ matches the actual observation $\boldsymbol{y}_i^\mathrm{o}$. 
In practice, however, there are always inevitable deviations from the truth state of the data represented by $\boldsymbol{d}_i = |\boldsymbol{y}_i^\mathrm{o} - \boldsymbol{y}_i |$. 

\paragraph{Data Assimilation (DA) schemes} Data assimilation techniques differ from data driving by explicitly solving a state-estimation problem. 
After a forecast step is calculated using Equation~(\ref{eq:data-driving}), $\boldsymbol{d}_i$ is calculated and used to produce an ``analysis'' state $\boldsymbol{x}^\mathrm{a}\left(t_i\right)$, improving self-consistency with the model and data realisation. In other words, the post-assimilation analysis state is expected to represent the data as best as possible while maintaining self-consistency with the model.
The formulation for data assimilation step is then of the form
\begin{eqnarray}\label{eq:data_assimilation_eqn}
    \boldsymbol{x}^\mathrm{a}\left(t_{i+1}\right) = \boldsymbol{x}^\mathrm{f}\left(t_{i+1}\right) +\boldsymbol{M}_i' \left[\boldsymbol{d}_i\right]
\end{eqnarray}
where $\boldsymbol{M}_i'$ is the dynamic operator for input $\boldsymbol{d}_i$. 
This differs from pure data-driving, where only the observation available at time $t_i$ enters the update, ignoring previous $\boldsymbol{d}_i$ information. 
In Section~\ref{sec:method_e_driving} we describe a corrected electric field that is required in our $\mathbf{E}$-driving strategy that formally makes it a data-assimilation scheme.


Adopting the framework above does more than standardise the terminology: it makes the existing mature body of techniques used in numerical weather prediction and geophysical modelling directly transferable, including ensemble forecasting \citep[e.g.][]{Abramowitz+2019}. 
Table~\ref{table:driving_methods_literature} summarises some contemporary data-driving strategies from the literature (some of which we will directly compare our results to in the remainder of this paper) and classifies them into the DD and DA categories. 


\begin{table*}[]
    \centering
    \begin{tabular}{|l|l|l|}
    \hline
    Author & Description & Classification \\ \hline 
    \hspace{-1cm} \begin{tabular}{l}
        M. C. Cheung \& \\ M. L. DeRosa \citeyearpar{cheung2012method}
    \end{tabular}

    &
    \hspace{-1cm} \begin{tabular}{l}
        Vector potential $\mathbf{A}$ driven by $\mathbf{E}$. Faraday $\mathbf{E}$ correction (time \\
        derivative of $B_z$ from observational/driving data used)
    \end{tabular}
    &
    \hspace{-1cm} \begin{tabular}{l}
        $\mathbf{E}$-driving (DD) \\ $\mathbf{A}$ is evolved 

    \end{tabular}
    \\ \hline
    
    \hspace{-1cm} \begin{tabular}{l}
        \citet{guo2019solar},\\
        \citet{guo2024magnetic},\\
        \citet{Guo2024FluxRope}
    \end{tabular}
    &
    \hspace{-1cm} \begin{tabular}{l}
        Observed data are imposed on the inner ghost layer of the bottom \\boundary, with zero-gradient extrapolation on the outer ghost layer.\\The normal component of the vector magnetic field on the outer\\ghost layer at the bottom boundary 
        are set to values that satisfy \\$\nabla\cdot\mathbf{B}=0$ to guarantee the solenoidal
        condition at the boundaries.
    \end{tabular}
    &
    \hspace{-1cm} \begin{tabular}{l}
        $\mathbf{B}$-driving (DD)\\
        with divergence \\ correction to \\ driving data
    \end{tabular}
    \\ \hline

    \hspace{-1cm} \begin{tabular}{l}
        \citet{he2020daremhd}\\
        \citet{Jiang2016}
    \end{tabular}
    &
    \hspace{-1cm} \begin{tabular}{l}
        AMR--CESE--MHD. The model uses a continuous input of the \\
        boundary state vector (calculated via a projected-characteristics\\
        method using wave-decomposition of the full MHD system).
    \end{tabular}
    &
    \hspace{-1cm} \begin{tabular}{l}
        $\mathbf{B}$-driving (DD) \\ Characteristic-based \\
        boundary driving
    \end{tabular}
    \\ \hline 

    Hoeksema et. al. \citeyearpar{Hoeksema2020CGEM}
    &
    \hspace{-1cm} \begin{tabular}{l}
        Vector potential $\mathbf{A}$ driven by the electric field.
    \end{tabular}
    &
    \hspace{-1cm} \begin{tabular}{l}
        $\mathbf{E}$-driving (DD) \\ $\mathbf{A}$ is evolved 
    \end{tabular}
    \\ \hline

    \citet{fan2024data}
    &
    \hspace{-1cm} \begin{tabular}{l}
        The horizontal electric field is imposed as a superposition of a\\PTD inductive field (reproducing the observed $\partial B_r/\partial t$), 
        a twisting\\potential derived from the observed $J_r$, and a random braiding\\field contribution. This is a constrained-transport code. 
    \end{tabular}
    &
    \hspace{-1cm} \begin{tabular}{l}
        $\mathbf{E}$-driving (DD) \\ 
    \end{tabular}
    \\ \hline

    \hspace{-1cm} \begin{tabular}{l}
        \citet{chen2023data}\\
        \citet{chen2026data} 
    \end{tabular}
    &
    \hspace{-1cm} \begin{tabular}{l}
        Electric-field driven, with no imposed mass flux. The 2023 method\\uses a Faraday electric-field correction 
        (time derivative of $B_z$ from \\observational/driving data used). The 2026 method solves directly\\for the $\mathbf{E}$-field via Faraday's law and dispenses with the correction.
    \end{tabular}
    &
    \hspace{-1cm} \begin{tabular}{l}
        $\mathbf{E}$-driving (DD)\\
        with corrected $\mathbf{E}$\\based on observations \\ only 
    \end{tabular}
    \\ \hline
    
    \citet{Liu2024DataDriven}

    &
    \hspace{-1cm} \begin{tabular}{l}
        Uses the electric field to update the magnetic induction equation on a\\
        finite-difference-based grid. No ghost cells are used. For observational\\
        data, the horizontal component is unreliable; Ohm's-law $E_z$ is\\
        retained, but the horizontal component is replaced with an electric-\\
        field inversion, split further into inductive and non-inductive \\
        components. DARE--MHD model.
    \end{tabular}
    &
    \hspace{-1cm} \begin{tabular}{l}
        Inverted $\mathbf{E}$-field (DD) \\
        driving
    \end{tabular}
    \\ \hline 
    
    \citet{hayashi2018mhd}
    &
    \hspace{-1cm} \begin{tabular}{l}
        Electric-field driven. The electric field is applied at three different\\ 
        $z$ heights: below, on, and above the boundary. Values are selected \\
        according to the prescribed boundary-driving procedure.
    \end{tabular}
    &
    \hspace{-1cm} \begin{tabular}{l}
        Multi-height\\
        $\mathbf{E}$-driving (DD) 
    \end{tabular}
    \\ \hline 

    \citet{Tarr2024MHDIII}
    &
    \hspace{-1cm} \begin{tabular}{l}
        This approach uses the method of characteristics to evolve a boundary\\
        state that is ``maximally'' close (optimisation sense) to the target state\\
        (e.g. ground truth data) and MHD-consistent. The MHD boundary \\
        state is interfaced to the LaRe3D code via an intermediate software, \\
        CHAR, which actually solves for the boundary state. LareXd $\rightarrow$ \\
        CHAR $\rightarrow$ \textit{Solve} $\rightarrow$ LareXd. The target MHD state in the boundary \\
        layer is calculated by interpolating between the current state and \\
        the next driving input state, making this data-assimilation. 
    \end{tabular}
    &
    \hspace{-1cm} \begin{tabular}{l}
        $\boldsymbol{U}$-driving (DA), i.e. \\the entire boundary 
        \\state $\left(\rho, \boldsymbol{v}, \mathbf{B}, E_\mathrm{int}\right)$, is\\driven. A characteristic-\\based solution of state $\boldsymbol{U}$\\is used.  
    \end{tabular}
    \\ \hline 

    \end{tabular} 
    \caption{Classification of the existing data-driving techniques in the literature  according to the terminology of \citet{ide1997unified}. Data driving (DD) and data assimilation (DA) can apply in any of the magnetic field ($\mathbf{B}$), electric field ($\mathbf{E}$), or full state ($\boldsymbol{U}$) driving methods. 
    }
    \label{table:driving_methods_literature}
\end{table*}

\section{Numerical Simulation Framework}\label{sec:methods}

We used Athena++ as the primary tool to simulate flux-emergence into the solar atmosphere. 
In this section, we discuss the critical concepts that are important for understanding data-driving strategies and our extended version of Athena++, now equipped with several $\mathbf{B}$- and $\mathbf{E}$-driving algorithms. 
The extensions to Athena++ allow the simulation of low $\beta$ plasmas required for solar modelling. 
A short presentation of the Athena++ code, as well as some details on the extensions to Athena++, is provided in Appendix~\ref{sec:method_athena}.

\subsection{Athena++ Solver Settings}
Athena++ is a finite-volume MHD solver which uses the ghost-cell method. 
We refer to regions of ghost-cells as ghost-zones. 
In a finite‑volume code, the ghost‑cell method enforces boundary conditions by populating fictitious cells just outside the computational domain with values prescribed or extrapolated so that the numerical fluxes (see Section \ref{sec:method_athena} for flux vector $\bf H$) computed at the domain boundary are consistent with the desired physical or mathematical boundary condition.
The following numerical techniques are used in the simulations presented in this work.
We use a three-dimensional, equally-spaced Cartesian grid, which is static (rather than adaptive) for simplicity. 
A second-order-accurate van Leer predictor-corrector scheme is used for time integration. 
A super-time-stepping scheme is used to handle the resistive terms, implemented with a second-order accurate Runge-Kutta-Legendre algorithm. 
The spatial reconstruction uses a piecewise Linear Method (PLM) with the van Leer limiter, requiring two ghost-cells around the simulation domain. 
In cases of high Alfv\'{e}n speeds, the Boris correction~\citep{boris1970nrl,gombosi2002semirelativistic} is used via a revised Harten-Lax-van Leer discontinuities (HLLD) flux solver \citep{miyoshi2005multi,matsumoto2019new}. 
In cases using an adiabatic equation of state, the dual-energy formalism \citep{bryan1995piecewise,takasao2022three} is used to enforce pressure positivity. 
The Boris correction and dual-energy formalism are enhancements to the standard Athena++ distribution by \citet{takasao2022three}.

\subsection{Side and Top Boundary Conditions}\label{sec:general_boundary_conditions}
In the simulations presented here the side boundaries ($x$ and $y$ dimensions) are treated using periodic boundary conditions. 
The top boundary is assumed to be closed for mass flux: 
the vertical velocity on the boundary face is enforced to be zero by an anti-symmetry condition on the ghost-cell $z$-velocity.
The horizontal velocities are treated identically to $z$-velocity for convenience, i.e. they are enforced to be zero on the upper $z$-boundary. In the simulations presented here the magnetic field remains very small at the top boundary (less than 2 mG) and there is no explicit viscosity modelled (precluding the formation of a viscous boundary layer), therefore these conditions do not significantly affect the simulation dynamics.

Modelling the gravitationally stratified  atmosphere requires appropriate values of density and pressure to be set in the ghost-cells adjacent to the top boundary.
In this work we specify the ghost-cell density and pressure using the equation of hydrostatic equilibrium with the domain value (upper $z$-boundary adjacent cell) as the reference condition. 
This results in a time varying Dirichlet condition at the top boundary for density and pressure. 

The magnetic field in the ghost-zones is typically defined in two ways. 
One method is to make a potential field extrapolation into the ghost-cells using the top domain cell $B_z$ values. 
The second is to assume a purely vertical  magnetic field (which implies a zero-gradient on the $B_z$ component into the ghost-cells). 
The latter is used in the current work for ease of simulation.

\subsection{Framing the data-driving implementation in Athena++}
We focus on simulating magnetic flux emergence through the lower $z$-boundary. 
We use a relatively simple treatment of the thermodynamic properties (density, velocity, and pressure): this  limitation of the present work reflects the limited information that is available from observational data. 
The density and pressure in the ghost-zones are set either as a constant, 
or according to a requirement of a hydrostatic equilibrium under gravity (described in Section~\ref{sec:appendix_atmosphere}). 

A crucial aspect of data driving is an interpolation procedure for taking the driving data and extracting desired quantities at different resolution (e.g.\ from observational resolution to simulation domain resolution) and grid locations (e.g., from cell-centred to staggered-grid data).
The present work uses bi-linear interpolation (e.g.\ see Section~\ref{sec:data_driving_comparison_toriumi}). 

In Section~\ref{sec:data driving} we explain conceptually the $\mathbf{B}$- and $\mathbf{E}$-driving strategies in Athena++. 
The discussion in Section~\ref{sec:data driving} uses ``driving data'' to refer to the external data  supplied to the data-driving algorithm. 
The driving data are used as inputs to the simulation, but they are not themselves part of either the physical simulation domain or the ghost zones. 
They may, however, be used to specify or constrain quantities in these regions.

In the discussions to follow, we distinguish three broad categories of boundary-driving strategies. 
In $\mathbf{B}$-driving, magnetic-field data are assigned directly to the ghost cells without solving for values consistent with the numerical boundary update; this method is therefore generally approximate (it is the conventional approach in the literature we argue is limiting). 
In $\mathbf{E}$-driving, the edge-centred boundary EMFs are prescribed directly, whereas ``update-consistent $\mathbf{B}$-driving'' denotes the hypothetical exact procedure in which the ghost-cell state is calculated by inverting the complete boundary-update algorithm to produce the required EMFs.




\section{The Data-Driving Algorithm} \label{sec:data driving}
\subsection{Data-driving as a boundary condition}\label{sec:frame_data_driving}


In essence, data driving is the formulation of appropriate boundary conditions for the MHD system. 
In a finite volume code, the fundamental goal of a boundary condition is to effect change in the simulation domain through the specification of the numerical fluxes on the boundary. 
In the case of the lower $z$-boundary, which is typically constructed to represent the photosphere, the numerical flux vector is given by 
\begin{equation}\label{eq:H_flux_vector}
{\bf H} =
\left[
\begin{array}{c}
\rho v_z \\[3pt]
\rho v_z v_x - B_z B_x \\[3pt]
\rho v_z v_y - B_z B_y \\[3pt]
\rho v_z^2 + P + B^2/2 - B_z^2 \\[3pt]
(E + P^\ast)v_z - ({\bf B}\cdot{\bf v})B_z \\[3pt]
B_x v_z - B_z v_x \\[3pt]
B_y v_z - B_z v_y \\[3pt]
0
\end{array}
\right]. 
\end{equation}
Boundary conditions are therefore set by controlling these boundary fluxes, which can be achieved either by prescribing ghost cell values that the stencil reads or by specifying the fluxes directly. 
Consequently, the key task of data driving reduces to forming the correct fluxes on the boundary of the domain. 
This is where the central challenge of data-driven solar MHD simulations arises: the solution of the MHD system requires that the numerical fluxes on the boundary be fully specified, but the correct numerical fluxes are not available from observations. 
As Equation~\ref{eq:H_flux_vector} for $\mathbf{H}$ makes explicit, the fluxes of mass, momentum, energy, and magnetic field depend on the full set of MHD variables. 
However, typically $\mathbf{B}$ and $\mathbf{v}$ are indirect measurements inferred from the observations via inversion methods that assume some model for the thermodynamic parameters that in turn introduces systematic errors \citep[e.g.][]{Borrero+2011}. 
Data-driving strategies are, in essence, the various attempts to bridge this gap and reconstruct the boundary fluxes that faithfully reproduce the desired coronal state from the limited information the observations provide.


\subsection{Magnetic Field Update Algorithm in Athena++} \label{sec:method_magnetic_update}
The magnetic field update in Athena++ is designed to be divergence-free by construction. 
The method of \citet{gardiner2005unsplit} was originally used
with the constrained-transport (CT) method of \citet{evans1988simulation}.
In Athena++, the magnetic field value defined on cell faces is evolved according to the integral form of the induction equation based on area rather than volume averages. 
Throughout this work, $\boldsymbol{\mathcal{E}}$ denotes the electric field. 
In the CT discretisation, its edge-centred representation is conventionally termed the electromotive force (EMF); we therefore use ``EMF'' specifically for edge-centred electric-field values used in the discrete induction update.
The magnetic field update is performed by the following equations (for a domain cell adjacent to the lower \(z\)-boundary, the underlined terms below depend on values in the ghost cells) 
\begin{subequations}\label{eq:Athena_B_update}
    \begin{align}\label{eq:magnetic_update_x}
        B^{n+1}_{x,i-1/2,j,k} &= B^{n}_{x,i-1/2,j,k} \, - \\
	&\frac{\delta t }{\delta  y} \left(  \mathcal{E}^{n+1/2}_{z,i-1/2,j-1/2,k} - \mathcal{E}^{n+1/2}_{z,i-1/2,j+1/2,k}\right) + \nonumber \\
&\frac{\delta t }{\delta  z} \left(  \,
\uline{\mathcal{E}^{n+1/2}_{y,i-1/2,j,k-1/2}}
- \mathcal{E}^{n+1/2}_{y,i-1/2,j,k+1/2}
\right) \nonumber
    \end{align}
	\begin{align}\label{eq:magnetic_update_y}
	   B^{n+1}_{y,i,j-1/2,k} &= B^{n}_{y,i,j-1/2,k} - \\
&\frac{\delta t }{\delta  x} \left(  \mathcal{E}^{n+1/2}_{z,i+1/2,j-1/2,k} - \mathcal{E}^{n+1/2}_{z,i-1/2,j-1/2,k}\right) +\nonumber \\
&\frac{\delta t }{\delta  z} \left(  \mathcal{E}^{n+1/2}_{x,i,j-1/2,k+1/2} 
- \uline{\mathcal{E}^{n+1/2}_{x,i,j-1/2,k-1/2}}\,\right) \nonumber
\end{align}
\begin{align}\label{eq:magnetic_update_z}
	B^{n+1}_{z,i,j,k-1/2} &= B^{n}_{z,i,j,k-1/2} - \\
&\frac{\delta t }{\delta  x}  \left(\, \uline{\mathcal{E}^{n+1/2}_{y,i+1/2,j,k-1/2}} - \uline{\mathcal{E}^{n+1/2}_{y,i-1/2,j,k-1/2}}
\, \right) + \nonumber \\
&\frac{\delta t }{\delta  y} \left( \,
\uline{\mathcal{E}^{n+1/2}_{x,i,j+1/2,k-1/2}} - 
\uline{\mathcal{E}^{n+1/2}_{x,i,j-1/2,k-1/2}}
\, \right), \nonumber
\end{align}
\end{subequations}
in which the superscript $n$ denotes a discrete time step, $\mathcal{E}_q$ is the $q$-component of the electric field (equivalent to EMF) in Cartesian coordinates,  $\delta t$ is a discrete time-step, and $\delta x$, $\delta y$, and $\delta z$ are discrete spatial steps in Cartesian coordinates. 
The subscripts $i$, $j$, $k$ are integer indices that denote
cell-centred spatial locations in the \(x\)-, \(y\)-, and \(z\)-directions,
respectively. 
The offsets $\pm 1/2$ denote quantities defined at cell edges, e.g. $(i,j+1/2,k+1/2)$, or cell faces, e.g. $(i,j+1/2,k)$.
\begin{figure*}
	\centering
	\tikzset{
		cube edge/.style={black},
		hidden edge/.style={black,dashed},
		vec/.style={->,>=Latex,thick},
		vec2/.style={->,>=Latex,thick,double},
		dot/.style={circle,fill=black,inner sep=1.5pt}
	}
	
	\begin{tikzpicture}[scale=3.5]
		
		\draw[vec] (-0.6,1.4)--+(0.42,0) node[right] {$X$};
		\draw[vec] (-0.6,1.4)--+(0.3,0.3) node[right] {$Y$};
		\draw[vec] (-0.6,1.4)--+(0,0.42) node[above] {$Z$};
		
		\begin{scope}
			
			\coordinate (A) at (0,0);
			\coordinate (B) at (1.2,0);
			\coordinate (C) at (1.2,1.2);
			\coordinate (D) at (0,1.2);
			\coordinate (E) at (0.4,0.4);
			\coordinate (F) at (1.6,0.4);
			\coordinate (G) at (1.6,1.6);
			\coordinate (H) at (0.4,1.6);
			
			\draw[cube edge] (A)--(B)--(C)--(D)--cycle;
			\draw[cube edge] (D)--(H)--(G)--(C)--cycle;
			\draw[cube edge] (B)--(F) (F)--(G) (C)--(G) (D)--(H);
			\draw[hidden edge] (A)--(E) (E)--(H) (E)--(F);
			
			\node[dot] at (0.7,0.7) {};
			\node at (0.7,0.6) [right,font=\large] {$U_{i,j,k}$};
			
			\node[dot] at (1.4,0.8) {};
			\draw[vec] (1.4,0.8)--+(0.4,0);
			\node[above,font=\large] at (1.8,0.8) {$B_{x,i^+,j,k}$};
			
			\node[dot] at (0.9,0.9) {};
			\draw[vec] (0.9,0.9)--+( 0.15,0.2);
			\node[right,font=\large] at (0.5,1.) {$B_{y,i,j^+,k}$};
			
			\node[dot] at (0.8,1.4) {};
			\draw[vec] (0.8,1.4)--+(0,0.5);
			\node[above,font=\large] at (0.8,1.9) {$B_{z,i,j,k^+}$};

		\end{scope}
		
		\begin{scope}[xshift=2.5cm]
			
			\coordinate (A) at (0,0);
			\coordinate (B) at (1.2,0);
			\coordinate (C) at (1.2,1.2);
			\coordinate (D) at (0,1.2);
			\coordinate (E) at (0.4,0.4);
			\coordinate (F) at (1.6,0.4);
			\coordinate (G) at (1.6,1.6);
			\coordinate (H) at (0.4,1.6);
			\draw[cube edge] (A)--(B)--(C)--(D)--cycle;
			\draw[cube edge] (D)--(H)--(G)--(C)--cycle;
			\draw[cube edge] (B)--(F) (F)--(G) (C)--(G) (D)--(H);
			\draw[hidden edge] (A)--(E) (E)--(H) (E)--(F);

			\node[dot] at (0.8,1.4) {};
			\draw[vec] (0.8,1.4)--+(0,0.28);
			\node[above, font=\large] at (0.85,1.65) {$H_{i,j,k^+}$};
			\node[dot] at (0.8,0.2) {};
			\draw[vec2] (0.8,0.2)--+(0,0.28);
			\node[below, font=\large] at (0.8,0.275) {$\boldsymbol{H_{i,j,k^-}}$};
			
%
			
			\node[dot] at (0.2,0.2) {};
			\draw[vec2] (0.2,0.2)--+(0.15,0.15);
			\node[left, font=\large] at (0.3,0.3) {$\boldsymbol{\mathcal{E}}^{y}_{i^-,k^-}$};
			
			\node[dot] at (1.4,0.2) {};
			\draw[vec2] (1.4,0.2)--+(0.15,0.15);
			\node[right, font=\large] at (1.5,0.15) {$\boldsymbol{\mathcal{E}}^{y}_{i^+,k^-}$};
			
			\node[dot] at (0.6,0) {};
			\draw[vec2] (0.6,0)--+(0.3,0);
			\node[below, font=\large] at (0.6,0) 
			{$\boldsymbol{\mathcal{E}}^{x}_{j^-,k^-}$};
			
			\node[dot] at (1,0.4) {};
			\draw[vec2] (1,0.4)--+(0.3,0);
			\node[right, font=\large] at (0.8,0.48) {$\boldsymbol{\mathcal{E}}^{x}_{j^+,k^-}$};
			
			\node[dot] at (1.4,1.4) {};
			\draw[vec] (1.4,1.4)--+(0.15,0.15);
			\node[right, font=\large] at (1.4,1.35) {$\mathcal{E}^{y}_{i^+,k^+}$};
			
			\node[dot] at (1.2,0.6) {};
			\draw[vec] (1.2,0.6)--+(0,0.3);
			\node[right, font=\large] at (1.2,0.6) {$\mathcal{E}^{z}_{i^+,j^-}$};
			
			\node[dot] at (1.6,0.9) {};
			\draw[vec] (1.6,0.9)--+(0,0.3);
			\node[right, font=\large] at (1.6,0.9) {$\mathcal{E}^{z}_{i^+,j^+}$};
			
		\end{scope}
	\end{tikzpicture}
	
	\caption{Athena++ staggered grid (left). The $Z$-direction numerical flux $\boldsymbol{H}$ and edge-centred electric field (equivalent to EMFs) values ($\boldsymbol{\mathcal{E}}$) at a driven boundary cell face (right). The terms labelled in boldface with double arrows are affected by ghost-cell states. The remaining terms are only affected by domain-cell values. For brevity, we use the short hand $i^{\pm} = i\pm1/2$ etc. and for $\boldsymbol{\mathcal{E}}$), the principle spatial index is dropped e.g. $\mathcal{E}^{x}_{i,j+\frac{1}{2},k+\frac{1}{2}} = \mathcal{E}^{x}_{j^+,k^+}$. 
	}
	\label{fig:boundary_driving_e_field}
\end{figure*}
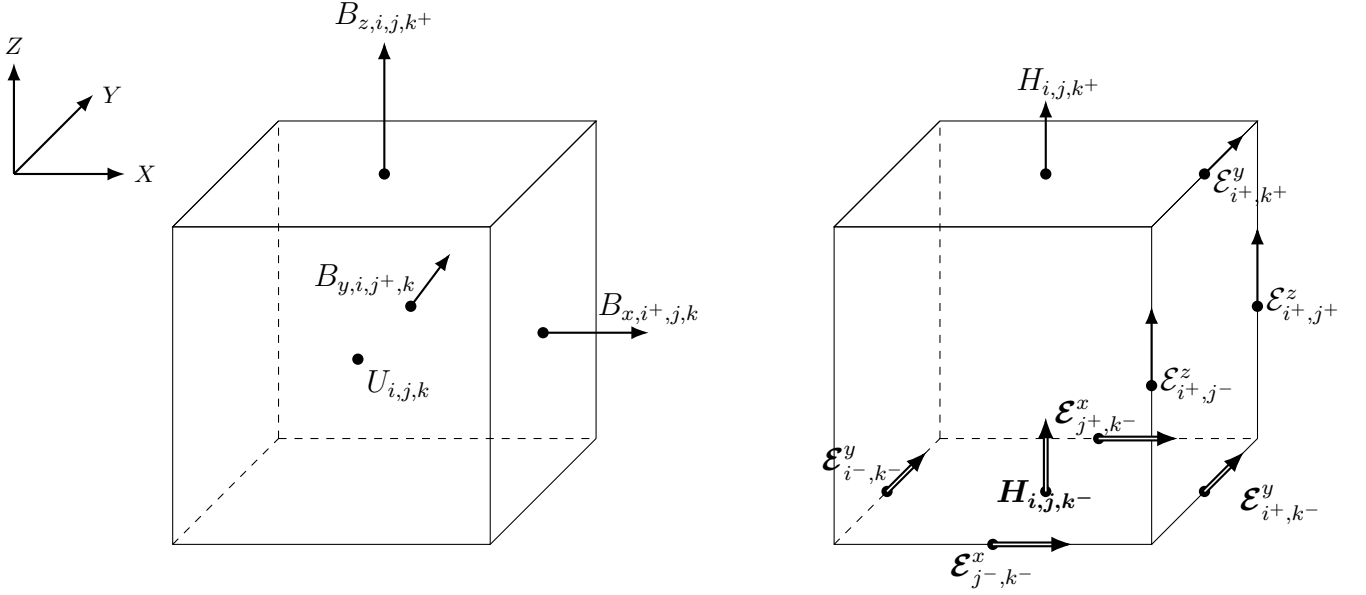
Figure~\ref{fig:boundary_driving_e_field}
illustrates how ghost-cell values influence the update scheme for a cell
adjacent to the domain boundary.

The electric field values are edge-centred in their relevant dimension, e.g., $\mathcal{E}_{x,i,j-1/2,k-1/2}$ is defined only on $x$-axis parallel edge centres. 
In CT terminology, the edge-centred electric-field quantity entering the discrete circulation in Faraday’s law is referred to as the EMF.
The algorithm for calculating the required electric field values uses (1) the cell-centred velocity and magnetic field values, and (2) the face-centred magnetic field fluxes. 
Cell-centred magnetic field values are calculated using second-order accurate averages of the adjacent face values, and the magnetic field fluxes are returned from the Riemann solver \citep{godunov1959finite}. 
A Riemann solver is a numerical method that computes the interface fluxes of a hyperbolic conservation law by approximating or exactly solving the local discontinuity problem defined by the left and right reconstructed states at a cell interface.

The quantities described above are combined to compute each edge-centred electric field.
For example, the $x$-component of the electric field at the edge
$x,i,j-1/2,k-1/2$ is given by
\begin{eqnarray}\label{eq:emf_calc}
	&&\mathcal{E}_{x,i,j-1/2,k-1/2}  =  \nonumber \\
	& & \frac{1}{4} \left( \mathcal{E}_{x,i,j-1/2,k} + \mathcal{E}_{x,i,j-1/2,k-1} \right. \nonumber \\
    & & \qquad + \left. \mathcal{E}_{x,i,j-1,k-1/2} + \mathcal{E}_{x,i,j,k-1/2} \right)
    \nonumber \\
	& & + \frac{\delta z}{8} \left( \frac{\partial  \mathcal{E}_{x,i,j-1/2,k-1/4}}{\partial z} - 
	\frac{\partial  \mathcal{E}_{x,i,j-1/2,k-3/4}}{\partial z}\right) \nonumber \\ 
	& & + \frac{\delta y}{8} \left( \frac{\partial  \mathcal{E}_{x,i,j-1/4,k-1/2}}{\partial y} - 
	\frac{\partial  \mathcal{E}_{x,i,j-3/4,k-1/2}}{\partial y}\right). 
\end{eqnarray}
The first term on the right-hand side of Equation~(\ref{eq:emf_calc})
is the arithmetic average of the surrounding face-centred electric field
values. 

In the Athena constrained-transport scheme, these face-centred electric
fields are obtained from the magnetic-field fluxes returned by the
Riemann solver. 
For example, at a $z$-interface, whose surface normal is
parallel to the unit vector $\boldsymbol{k}$, the fluxes of the
interface-parallel magnetic-field components $B_x$ and $B_y$ are
$-E_y$ and $E_x$, respectively, while the flux of $B_z$ is zero.

These face-centred electric field values therefore depend on the Riemann-solver solution at the corresponding cell interface. 
In the present work, the HLLD Riemann solver \citep{miyoshi2005multi}, or its isothermal variant \citep{mignone2007simple}, reconstructs the appropriate interface state from the left and right states adjacent to the interface. 
The remaining terms in Equation~(\ref{eq:emf_calc}) apply transverse gradient corrections in the $z$- and $y$-directions. 
These corrections are constructed from weighted combinations of the Riemann-solver electric fields and the neighbouring cell-centred electric field values surrounding the edge.
These terms in Equation~(\ref{eq:emf_calc}) arise from the CT EMF reconstruction, formulated by \citet{stone2008athena} to maintain multidimensional up-winding and consistency between the finite-volume and CT updates.

In summary, a data-driving boundary condition must prescribe the calculation of $\mathcal{E}_{x}$ and $\mathcal{E}_{y}$ at the boundary. 
These edge-centred electric fields are obtained from a non-local stencil involving Riemann-solver electric fields and neighbouring cell-centred electric field values. 
To control the Riemann-solver contribution, one must control the reconstructed left and right interface states, which depend on cell-centred values in both the physical domain and the ghost-cell zones. 
In addition, the ghost-cell values of $\mathbf{B}$ and $\mathbf{v}$ directly enter the calculation of the cell-centred electric field used in the CT EMF construction.
Consequently, determining the ghost-cell values required to impose a prescribed update of the lower \(z\)-boundary magnetic field is non-trivial.

\subsection{$\mathbf{B}$-driving---Magnetic Field Data Driving}\label{sec:method_b_driving}
In the $\mathbf{B}$-driving strategies considered here, magnetic-field data, e.g. ground-truth simulation data, are assigned directly to the ghost cells without inverting the numerical boundary-update algorithm. 
These prescriptions therefore constrain the driven boundary evolution only approximately.
On a Cartesian grid, as considered here, the driven boundary is the lower $z$-boundary.
In Athena++, the ghost cells are manipulated to impose the desired boundary state, from which the Riemann solver computes the numerical fluxes at the driven boundary.
Although these boundary fluxes can be specified directly from the driving data, this bypasses the usual flux construction and can break consistency between the conserved variables, primitive variables, and induction update.
Consequently, direct flux prescription can generate spurious MHD states, numerical instability, and violations of the solenoidal constraint ($\nabla\cdot\mathbf{B}=0$).
Likewise, it is possible to overwrite domain values of the magnetic field (and other variables) but this suffers from the same drawbacks.
Using the ghost-cell method to accurately control the boundary driving requires careful selection of ghost-cell values, values that depend on the magnetic field update algorithm (see Section~\ref{sec:method_magnetic_update} for a detailed explanation of the Athena++ update). 



The simplest conventional $\mathbf{B}$-driving strategy directly assigns the driving magnetic field data to the ghost cells. 
Similarly, the ghost-zone velocities are directly assigned from driving data---they should be specified consistently with the magnetic field so that Ohm's law recovers the electric field associated with the desired constrained-transport update of the magnetic field.
A variation of this method used in non-constrained-transport codes is to modify the ghost-zone magnetic field values to enforce the solenoidal constraint for the magnetic field. 
For example \cite{guo2019solar} set the first ghost-zone layer to the driving-data values and then adjusted $B_z$ in the second ghost-zone layer, using a second-order reconstruction at the boundary, to enforce $\nabla\cdot\mathbf{B}=0$. 

The implementation above, while providing an approximate data-driving strategy, does not solely control the magnetic field evolution on the boundary. 
The ghost-zone values also affect the Riemann-solver contribution at the driven boundary. 
In particular, they enter the reconstruction of the left and right interface states supplied to the Riemann solver. 
This reconstruction depends on both ghost-zone and domain-zone values, with a stencil determined by the chosen reconstruction method. 
The Riemann solver then uses these reconstructed states to compute the fluxes required for the finite-volume update of the domain, including the update adjacent to the driven boundary.

Alternatively, rather than filling the ghost-zones directly using the driving data, ghost-zone values can be set by requiring that the driving-data values are matched at the interface between the ghost-zone and physical domain.
For example, a linear extrapolation using the first domain cell (adjacent to the boundary) and the interface value (matching the driving data) can be used to calculate ghost-zone values. 
The interface value that is subsequently reconstructed within the Athena++ update algorithm and provided as input to the Riemann solver will then match the driving data.
However, any cell-centred electric-field values, which contribute to the subsequent edge-centred EMF construction (see Section~\ref{sec:method_magnetic_update}) are not guaranteed to be physically meaningful or consistent with the desired update. 
This approach was implemented in Athena++ and found to be numerically unstable and inaccurate for data driving. 

Consider Equations~(\ref{eq:magnetic_update_x})-(\ref{eq:magnetic_update_z}), which describe the magnetic field update on the lower $z$-boundary.
In general, reproducing a desired magnetic field cannot be achieved by specifying ghost-cell values of magnetic fields from observational (or analytical data) alone. 
To control Athena++'s magnetic field update exactly, we can either (1) prescribe the required edge-centred EMFs directly, as in $\mathbf{E}$-driving (Section~\ref{sec:method_e_driving}) or (2) prescribe the ghost-cell state that produces those EMFs through the complete numerical update, hereafter update-consistent $\mathbf{B}$-driving. 

In update-consistent $\mathbf{B}$-driving, option (2) above, the strictly correct approach is to essentially invert the update algorithm and solve for the ghost-cell values required to produce the desired properties on the interface. 
However, implementation of such an inversion is highly involved:
one would first specify the desired magnetic-field state, or equivalently the desired boundary update, and use the constrained-transport induction equation (Equations \ref{eq:magnetic_update_x}-\ref{eq:magnetic_update_z}) to determine the edge-centred EMFs required to produce it.
The ghost-cell variables must then be chosen so that after (i) reconstruction, (ii) the Riemann solve, and (iii) the subsequent CT EMF construction (Equation \ref{eq:emf_calc}), the computed edge-centred EMFs match these required values.
This defines an inverse problem for the ghost-zone magnetic field and velocity,
with the interior-domain state treated as fixed in a single time step. 
The Riemann-solver inversion itself is mathematically challenging; see \citet{matsumoto2019new} for an example of the HLLD Riemann solver equations that would have to be inverted. 
The reconstruction is tractable; see \citet{van1977towards} for an example of the equations to invert in a Van Leer limiter reconstruction method. 
In contrast, the $\mathbf{E}$-driving procedure directly prescribes the required edge-centred boundary EMFs and therefore avoids the inverse problem required for update-consistent B-driving.

This logic {\color{red}generally applies to magnetic-field update algorithms in which the boundary update is determined through numerical fluxes or discrete derivative stencils involving ghost-cell and physical domain values; it is not specific constrained-transport-based algorithms}. 
E.g. in a finite-volume scheme such as Athena++, the boundary update depends on the reconstructed interface states, Riemann solver fluxes, and the CT EMFs, e.g. in a finite-difference scheme such as MURaM, the boundary update depends on the discrete derivative stencil used to evaluate the induction equation and the associated electric field reconstruction. Consequently, matching the ghost-cell values to driving data does not generally imply matching the interface state or its time evolution to the driving data.

Conventional implementations of $\mathbf{B}$-driving that assign ghost values directly from driving data without accounting for the relevant boundary update algorithm are not correctly controlling the boundary condition numerically. These methods are generally under-constraining the boundary condition, and we refer to these methods as ``approximate data-driving'' strategies. In some situations, the approximate data-driving approach will produce an effective boundary condition but in others (as we show in Section~\ref{sec:data_driving_comparison_fan}) they deviate with significant errors. 
An analogy would be to set a zero-velocity condition on a boundary by imposing zero ghost-cell values. 
The correct approach is to calculate the appropriate ghost-cell velocities at each time step by using the domain velocities and the code's spatial reconstruction scheme (in the simplest case setting the ghost-cell value to the negative of the domain value assuming a linear spatial reconstruction).


Many other $\mathbf{B}$-driving approaches exist in the literature, see Table~\ref{table:driving_methods_literature}, but they tend to be approximate in that they do not provide direct justification for choices based on the magnetic field update algorithm. 
Additionally, there are the issues of introducing divergence errors into the simulation domain, or ad-hoc corrections which further deviate from the true driving data state. 
These problems arise from limited observational data (there is limited knowledge of magnetic field variation with depth at the photosphere) and the complexity of developing first principles boundary conditions.
\citet{Tarr2024MHDIII} have developed a method of characteristic boundary condition for MHD which addresses the issues discussed. 
The method uses a separate characteristic-based MHD solver (CHAR \citep{Tarr2024MHDIII}) to determine a MHD consistent boundary state that is maximally close to the target state, i.e.\ the driving data/observation, although the evolved state can deviate significantly from the driving data in some circumstances \citep{Tarr2024MHDIII}.

We use two $\mathbf{B}$-driving approaches in this work. 
The first is $\boldsymbol{v}$-$\mathbf{B}$-driving, used in Section~\ref{sec:data_driving_comparison_fan}, where spatially-varying driving data ($f(x,y,z)$) is available to assign the face-centred magnetic field components and cell-centred velocity components throughout the ghost zone at the lower boundary. 
The second approach, used in Section~\ref{sec:data_driving_comparison_toriumi}, uses only driving data from a single vertical height, i.e., data are of the form $f(x,y)$. 
A linear extrapolation (along the $z$-axis) of velocity into the ghost-zone cell-centres is used to match velocity driving-data on the boundary. 
For the magnetic field, the $x$-$y$-varying driving-data is interpolated to face-centred $x$ and $y$ positions for the magnetic field components and applied to both layers of ghost-cells i.e., neglecting $z$-spatial variation. 
A variation of this strategy is also considered, where the outer ghost-cell $B_z$ is set to a value that enforces the divergence constraint $\nabla\cdot\mathbf{B}=0$ in the ghost zone. 

\subsection{$\mathbf{E}$-driving --- Electric Field Data Driving (Data Assimilation)}\label{sec:method_e_driving}

$\mathbf{E}$-driving in Athena++ is implemented by directly prescribing the edge-centred EMFs, $\boldsymbol{\mathcal{E}}$, representing the lower $z$-boundary electric field in the CT update.
Specifically, for all $i$ and $j$ across the data-driven boundary (where $k_l$ is the first domain cell in $z$): 
\begin{eqnarray}
\mathcal{E}^{vB}_{x,i,j+1/2,k_l-1/2} &=& -(v_y B_z - v_z B_y)^{\rm obs} \qquad \textrm{and}\\ 
\mathcal{E}^{vB}_{y,i-1/2,j,k_l-1/2} &=& (v_x B_z - v_z B_x)^{\rm obs}.  
\end{eqnarray}
The $\mathcal{E}^{vB}$ values are calculated from the driving-data magnetic field and velocity using the ideal Ohm's law (and additions to  Ohm's law can be made to account for non-ideal MHD terms, e.g.,~resistivity). 
In this way, the instantaneous evolution of $B_z$ on the driven boundary can be made to reproduce exactly (within numerical and interpolation error bounds) the driving-data. 
The values of $B_x$ and $B_y$ in the first domain cell (evolved at face-centres) are influenced by the specified boundary values of $\mathcal{E}_{x}$ and $\mathcal{E}_{y}$.
However, they are evolved self-consistently by the update algorithm inside the domain, and therefore need not match the driving data.
Note that specifying $\mathcal{E}_{x}$ and $\mathcal{E}_{y}$ on the driven boundary is equivalent to setting the fluxes of $B_x$ and $B_y$ on the driven boundary (
Equation~(\ref{eq:H_flux_vector})).

The instantaneous time derivative of $B_z$ given by the driving electric field ($\boldsymbol{\mathcal{E}}^{vB}$)
\begin{eqnarray}
	-\frac{\partial B_z}{\partial t} =  \left(\nabla \times \boldsymbol{\mathcal{E}}^{\;vB}\right)\cdot \hat{z}.
\end{eqnarray}
does not guarantee the time-averaged vertical component of Faraday's equation (in regard to driving data cadence) because
\begin{eqnarray}
	-\frac{\Delta B^{\rm obs}_z}{\Delta t} \neq  \left(\nabla \times \boldsymbol{\mathcal{E}}^{\;vB}\right)\cdot \hat{z}.
\end{eqnarray}
To satisfy Faraday's equation for the time evolution of $B_z$, a correction electric field can be calculated following the method of \cite{cheung2012method}, by solving  
\begin{eqnarray}\label{eq_cheung_electric}
	\left(\nabla \times \boldsymbol{\mathcal{E}}^{C}\right)\cdot \hat{z} &=& -\frac{\Delta B^{\rm obs}_z}{\Delta t} \label{eq:e_corrected_driving}\quad \textrm{and}\\ 
	\nabla_h \cdot \boldsymbol{\mathcal{E}}^{C} &=& D\left(x,y\right).\label{eq:e_corrected_driving_divE}
\end{eqnarray}
However, this correction to the electric field did not improve our results with Athena++ i.e. the quantitative deviation of the evolved $B_z$ on the boundary to driving data. 
In our implementation, an important additional source of error is interpolation of driving data to (i) Athena's staggered grid (from the cell-centred driving data) and (ii) different resolutions. 
There are additional error contributions from numerical diffusion and the limited cadence of sampling (i.e.\ the driving-data cadence of $\Delta t = 1$ non-dimensional units compared to the much smaller internal dynamic hyperbolic and parabolic time-step constraints).

We instead calculate a correction electric field using the residual of the realised state i.e. the deviation of the evolved boundary value of $B_z$ from the driving data.


In general the driving data is available at some discrete set of times $t=\tilde{t}^i, i=0\ldots i_{\rm max}$. Let us denote the driving data at these times by $\tilde{\bf B}^i=\tilde{\bf B}(\tilde{t^i})$.
Given the smaller simulation timestep, a given driving electric field is then applied for multiple simulation timesteps i.e., the corresponding boundary EMFs are held constant over multiple simulation time steps.
Then, for example, starting from $t=\tilde{t}^0=0$, ideally, when the simulation reaches $t=\tilde{t}^1$ the boundary magnetic field will match the driving data, $\tilde{\bf B}^1$. However, in general this is not the case, and for this reason we calculate a correction term to account for the discrepancy, as follows.

In practice, the simulation times will never exactly match the driving times, but consider what happens the first time that the simulation time exceeds the time of the $i^{\rm th}$ driving data snapshot. Denote this time as $t=T^i$. Since $T^i-\tilde{t}^i \ll\tilde{t}^i-\tilde{t}^{i-1}$, the accumulated discrepancy in the magnetic field (or residual) is approximately
\begin{eqnarray}
        \Delta B_z = \tilde{B}_z(t=\tilde{t}^i) - B_z(t=T^i), 
\label{eq:delta-Bz}
\end{eqnarray}
where the first term is the ``target'' value of $B_z$ from the driving data and the second is the realised state, both on the $z=0$ boundary.

Having evaluated $\Delta B_z$, we then calculate and apply a correction electric field
$\boldsymbol{\mathcal{E}}^{\textrm{C}}$ to retrospectively correct (cancel out) the deviation accumulated between $\tilde{t}^{i-1}$ and $\tilde{t}^{i}$. 
In this way, the correction $\boldsymbol{\mathcal{E}}^{\textrm{C}}$ works with data-driving emf $\boldsymbol{\mathcal{E}}^{\textrm{vB}}$ by the superposition $\boldsymbol{\mathcal{E}} = \boldsymbol{\mathcal{E}}^{\textrm{C}} +  \boldsymbol{\mathcal{E}}^{\textrm{vB}}$. 
In general, the correction is calculated as
\begin{eqnarray}
	\left.\left(\nabla \times \boldsymbol{\mathcal{E}}^{\textrm{C}}\right)\cdot\hat{z}\right|_{T^{i}<t<T^{i+1}}   &=& -\frac{\Delta B_z}{\tilde{t}^{i+1}-T^i } ,\label{eq:e_corrected_assimilation_curl} 
\end{eqnarray}
with $\Delta B_z$ given by Equation~(\ref{eq:delta-Bz}).

In the notation of Section~\ref{sec:driving_vs_assimilation}, the deviation {$\Delta B_z$} is the observational residual, $d_i$, and the superposed correction $\boldsymbol{\mathcal{E}}^{\textrm{C}}$ represents the analysis step contribution in Equation~(\ref{eq:data_assimilation_eqn}): the boundary update incorporates the tracked forecast-observation deviation rather than the observation alone.
$\boldsymbol{\mathcal{E}}^{\textrm{C}}$ is calculated using a Fourier space solution of Equations~(\ref{eq:e_corrected_assimilation_curl}) and~(\ref{eq:e_corrected_driving_divE}). 
By contrast, the right hand side of Equation~(\ref{eq:e_corrected_driving_divE}) accounts for error in the electric field divergence on the lower boundary. 
This quantity implicitly controls the injection of magnetic free energy and magnetic helicity. 
The discrete numerical function $D\left(x,y\right)$ is obtained by comparison of the discrete divergence of the ideal Ohm's law $\mathbf{E}_h$ computed from the 
original driving data and computed from the staggered-grid interpolated driving data---the latter is the electric field actually used in the $\mathbf{E}$-driving within Athena. 
The ad-hoc corrections, introduced to mitigate interpolation errors in the evolution of the $B_z$ component, are not guaranteed to improve the evolution of the horizontal components. Although both the horizontal and $B_z$ magnetic field components are influenced by the same errors (via $\mathcal{E}_x$ and $\mathcal{E}_y$), the correction may produce a deviation from the true driving state in the horizontal components.

In the Athena++ $\mathbf{E}$ driving, ghost-cell $\mathbf{B}$ values affect the energy flux through the lower boundary, which is essential for preserving pressure positivity in boundary adjacent cells.
Only the prescribed $\boldsymbol{\mathcal{E}}$ is used to advance the magnetic field at the driven boundary, but $\mathbf{B}$ values in ghost-cells must still be supplied. 
Two approximate treatments are adopted: either prescribing $\mathbf{B}$ values from the driving data or imposing a symmetry condition at the lower boundary. 
Both approaches yield satisfactory results, with no evidence of negative pressures arising from inadequate energy flux. 

\subsection{Lower $z$-boundary conditions on plasma variables}
Several different velocity boundary conditions have been experimented with in Athena++ to investigate the effect on data driving. 
We set velocity boundary conditions by assigning ghost-zone values from driving data (Dirchlet condition) or using generic boundary conditions set from the domain (e.g. as described in Section~\ref{sec:method_b_driving}. 
For $\mathbf{B}$-driving, setting Dirichlet conditions on velocity from driving data is the most appropriate boundary treatment due to the Athena++ update algorithm that relies on ghost-cell calculations of cell-centred electric fields from $\mathbf{B}$ and $\boldsymbol{v}$. 
Appendix~\ref{sec:Bdriving_innapropriate} shows the problems with zero mass-flux and zero-gradient conditions for $\mathbf{B}$-driving. 
The Athena++ $\mathbf{E}$-driving strategies can use any of the three velocity boundary conditions described without degrading magnetic field evolution significantly (this is shown in Section~\ref{sec:data_driving_comparison_toriumi}). 
Using observational or ground truth data for the velocity boundary condition is preferable, as it is more physically meaningful than the zero mass-flux and zero-gradient velocity boundary conditions. 
Some authors make a distinction between outflow and inflow boundary conditions in data driving, and while not used here it is discussed by \citet{vogler2003three}. 
Depending on the simulation case, however, it may be advantageous to use zero-mass flux or zero-gradient velocity boundary conditions for $\mathbf{E}$-driving---only the incoming characteristic modes should be prescribed on a boundary and over-specifying the boundary can cause numerical instability.

Density and pressure boundary conditions are treated together, taking into account the gravitationally stratified atmosphere to maintain a quasi-hydrostatic equilibrium on the driven boundary with non-zero velocities. 
When the simulation is initialised, the atmospheric profile of density and pressure is calculated from hydrostatic equilibrium and applied to the domain and ghost zones. 
The values that are initialised in the ghost zones are maintained throughout the simulation as a Dirichlet boundary condition. 
This choice is made in the absence of any knowledge of the 
density and pressure below the driving boundary, and to avoid a negative acceleration due to gravity on the boundary. 
Note that solving an atmospheric profile with the lower $z$-boundary adjacent cell as a reference point can be erroneous in the case of upflows. 
When there is compression due to an upflow, the pressure and density increase, requiring a larger density and pressure below to support the atmosphere above. 
A positive feedback loop ensues, producing increasingly larger and non-physical density and pressures being injected at the bottom boundary.

\section{Simulated Emergence of an Idealised Toroidal Flux-tube}\label{sec:fangibson}
\subsection{Idealised Toroidal Flux-tube: Setup}
\begin{figure}
	\centering
	\plotone{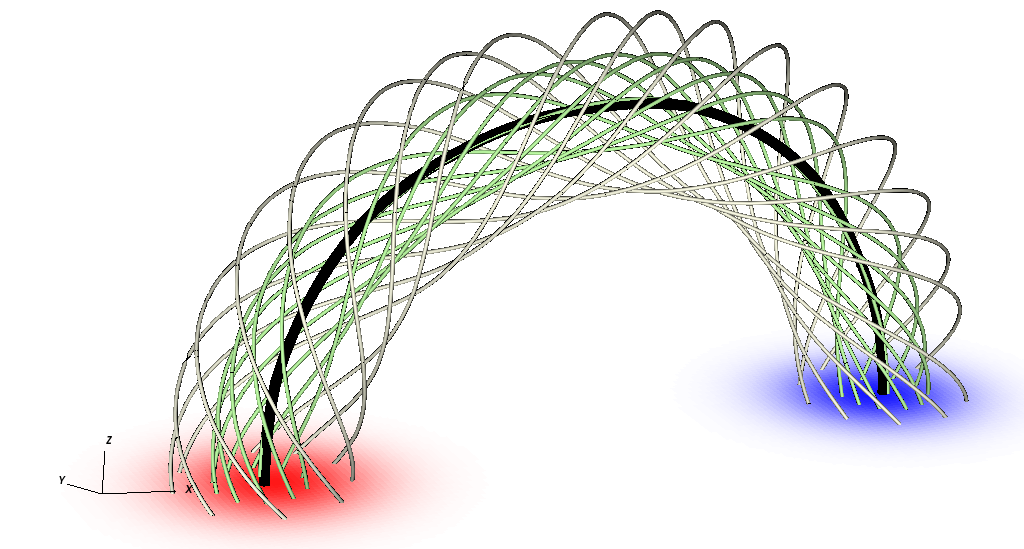}
	\caption{An example of the toroidal flux-tube defined by the model of \citet{fan2003emergence} that is simulated in this work (using $q=-1$ twist). \label{fig:sun_flux_tube}}
\end{figure}

To validate the implementation of the data driving in Athena++ we first reproduce an augmented version the flux-emergence problem designed by \citet{fan2003emergence}.
The flux-tube magnetic field is constructed as a torus whose axis lies in the $y=0$ plane, with major radius $R$, and with a specified twist (see Figure \ref{fig:sun_flux_tube}). 
The  flux tube rises with constant vertical speed $v_0$ into the domain. 
\citet{sun2021non} give the expression for the magnetic field components in Cartesian coordinates.
Figure \ref{fig:sun_flux_tube} shows an example of the model flux-tube with negative twist. 

In the present study, the following reference values are used:
\begin{eqnarray*}
	\rho_0 =& 1\times10^{-7} \, \textrm{g/cm\textsuperscript{3}}\text{, }\,  L_0 &= 1\times10^{4} \quad \textrm{cm}
\end{eqnarray*}

which then parameterise the flux-tube with
\begin{eqnarray*}
    v_A &=1\times10^{4} \, \textrm{cm/s} \quad\quad B_0 &= \sqrt{4\pi\rho_0v_A^2}\\
	v_0 =& 0.0125 v_A\text{, } \quad\quad w &= 0.05L_0  \\
	y_{-} =&  0.4L_0 \text{, } \quad\quad y_{+} &= -0.4L_0  \\
	R =& 0.375L_0   \text{, } \quad\quad q_0 &= -1 \\
	a_0 =& 0.1L_0 \text{, and } \quad\quad B_t &= 9B_0. 
\end{eqnarray*}
Here $w$ 
is the characteristic distance from the tube axis, $B_t$ is the characteristic field strength, $q_0$ sets the twist, and $a_0$ sets the characteristic length scale for the minor axis of the torus. 
The magnetic field lines wind about the axis at a rate of $q_0/a_0$ per unit length. 
The magnetic field of the flux-tube is truncated to zero outside the toroidal flux surface defined by minor radius $a_0$. 
We use a negative twist, similar to the simulation by \citet{fan2024data}. 
Our simulation here differs from the original by removing the background potential magnetic field arcade and using a higher beta plasma. 

The analytical definition of the magnetic field within the flux-tube is used to calculate the relative magnetic helicity in the simulation domain ($H_R$), using the method of  \citet{valori2012comparing} and total unsigned flux on the $z=0$ boundary ($\Phi$, surface integral of $B_z$). 
This provides the ideal $H_R(t)$ and $\Phi(t)$ reference values for comparison with the simulation derived $H_R(t)$ and $\Phi(t)$. 
$H_R$ and $\Phi$ are calculated for the entire simulation time, before and after the rise end time. 
The rise end time, referred to as ``anchoring'', is when the flux tube vertical motion ceases and it is kept stationary at the boundary.
Using the flux-tube model parameters described above, the flux emergence is reproduced in the Athena++ code with conventional ghost-cell $\mathbf{B}$-driving and $\mathbf{E}$-driving; update-consistent $\mathbf{B}$-driving is not implemented.

In regard to the boundary conditions, the complete spatial and temporal variation of the velocity and magnetic field vectors is known in this test case. 
This means in the $\mathbf{B}$-driving case, ghost-cell values can be specified directly without any assumptions or extrapolations being required.
Likewise, edge-centred boundary EMFs along the driven boundary (lower $z$-boundary) can be calculated from the analytical electric field obtained using the ideal Ohm’s law---the correction electric field discussed in Section~\ref{sec:method_e_driving} is not used or required as there is no interpolation of data used in this data driving problem.
The background plasma is assumed uniform isothermal.
Given the assumed uniform vertical rise of the flux tube we set $\boldsymbol{v}=\{0,\;0,\;v_0\}$ throughout the ghost zones during the rise, and after the rise ends, $v_0$ is set to zero.

\subsection{Idealised Toroidal Flux-tube: Comparison of Data Driving Strategies}\label{sec:data_driving_comparison_fan}
We have simulated the flux emergence using both the $\mathbf{E}$-driving and $\mathbf{B}$-driving methods described above. 
We calculate for each time step the relative magnetic helicity $H_R$ using the method of \citet{valori2012comparing}, using a potential field extrapolation from the NDSM code (\url{https://github.com/sag2021/ndsm}) with the normal components of the simulation $\mathbf{B}$ taken as the boundary conditions on all six faces. 
The flux $\Phi$ at $z=0$ is calculated using the face-centred $B_z$ values evolved in the simulations. 

\begin{figure}
    \plotone{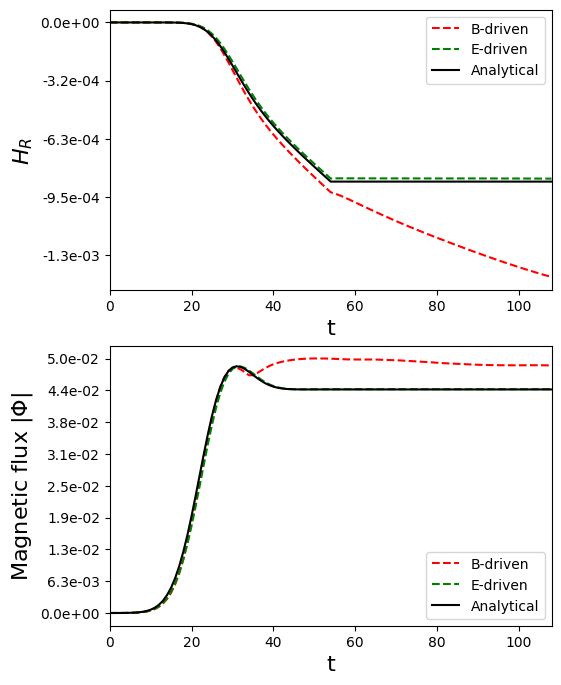}
	\caption{Relative magnetic helicity (top, units of $B_0^2L_0^4$) and total unsigned flux on the driven boundary (bottom, units of $B_0L_0^2$), for the emergence of an idealised flux tube (Fig.~\ref{fig:sun_flux_tube}). The calculated time series represent results for the analytical model (Analytical) and the simulation results from the $\mathbf{E}$-driven strategy, and $\mathbf{B}$-driven strategy. \label{fig:fan_gibson_helicity_flux}}
\end{figure}

The resulting time series for $H_R$ and $\Phi$ are shown in Figure \ref{fig:fan_gibson_helicity_flux}.
The analytical result represents a highly idealised flux emergence that does not include any deformation or work against the surrounding plasma.
Consequently, in a full MHD evolution we would expect some deviation in the thermodynamic evolution and local magnetic field distribution (e.g., the flux tube height) while global magnetic field properties like $H_R$ and $\Phi$ should closely match the analytical model. 
Both time series in Figure~\ref{fig:fan_gibson_helicity_flux} indicate that the $\mathbf{E}$-driving simulation provides the closest reproduction of the analytical result. 
$\mathbf{B}$-driving here deviates in $H_R$ modestly during rise time ($t\leq 54$) and then significantly over injects $H_R$, $\Phi$ follows a similarly poor reproduction of the analytical result. 
The increase in $\Phi$ (Figure~\ref{fig:fan_gibson_helicity_flux} B-driven series) at $t\approx 34$ is due to the spurious magnetic field evolved between the main flux tube polarities after they have separated. This effect is absent in the $\mathbf{E}$-driven case because the boundary magnetic field is more accurately evolved using the EMFs.

Figure 
\begin{figure*}
	\centering
	\begin{subfigure}[b]{.45\linewidth}
		\includegraphics[width=\linewidth]{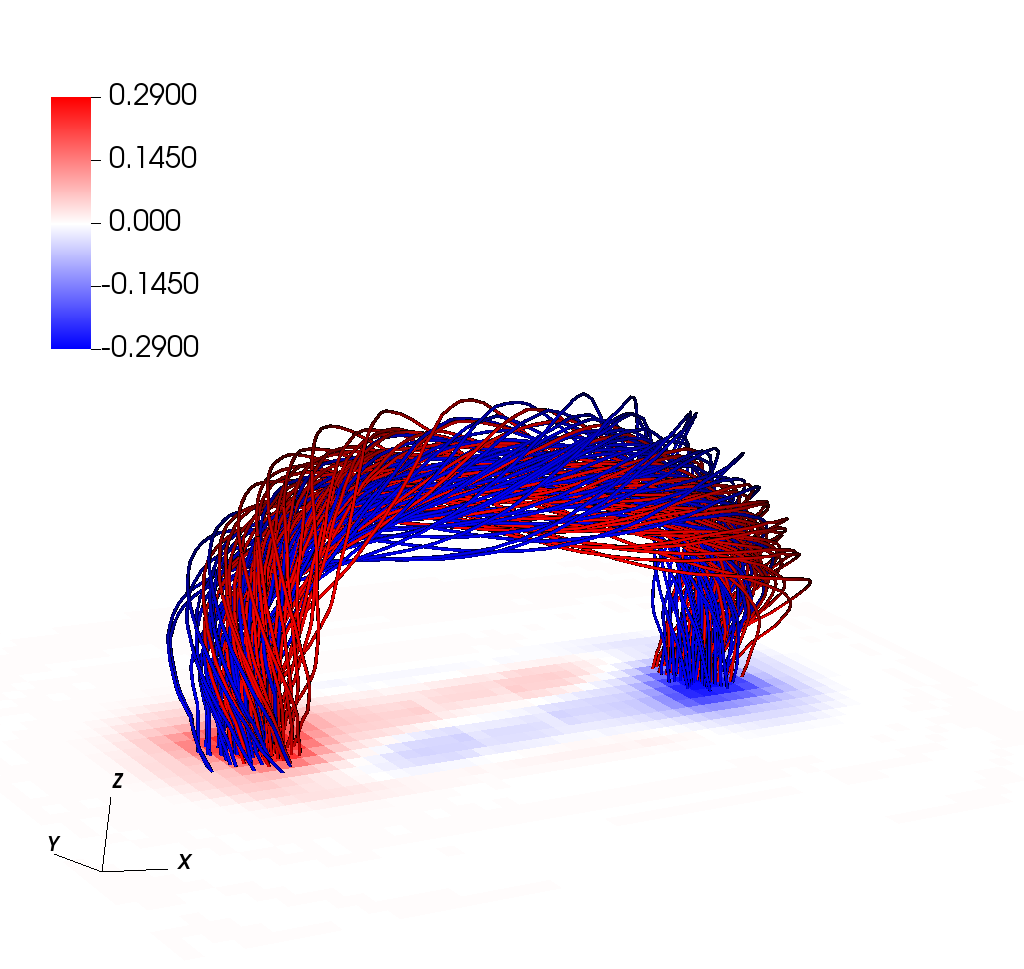}
		\caption{End of the rise time via $\mathbf{B}$-driving. }\label{fig:FG_rise_end_B_driving}
	\end{subfigure}
	\begin{subfigure}[b]{.45\linewidth}
		\includegraphics[width=\linewidth]{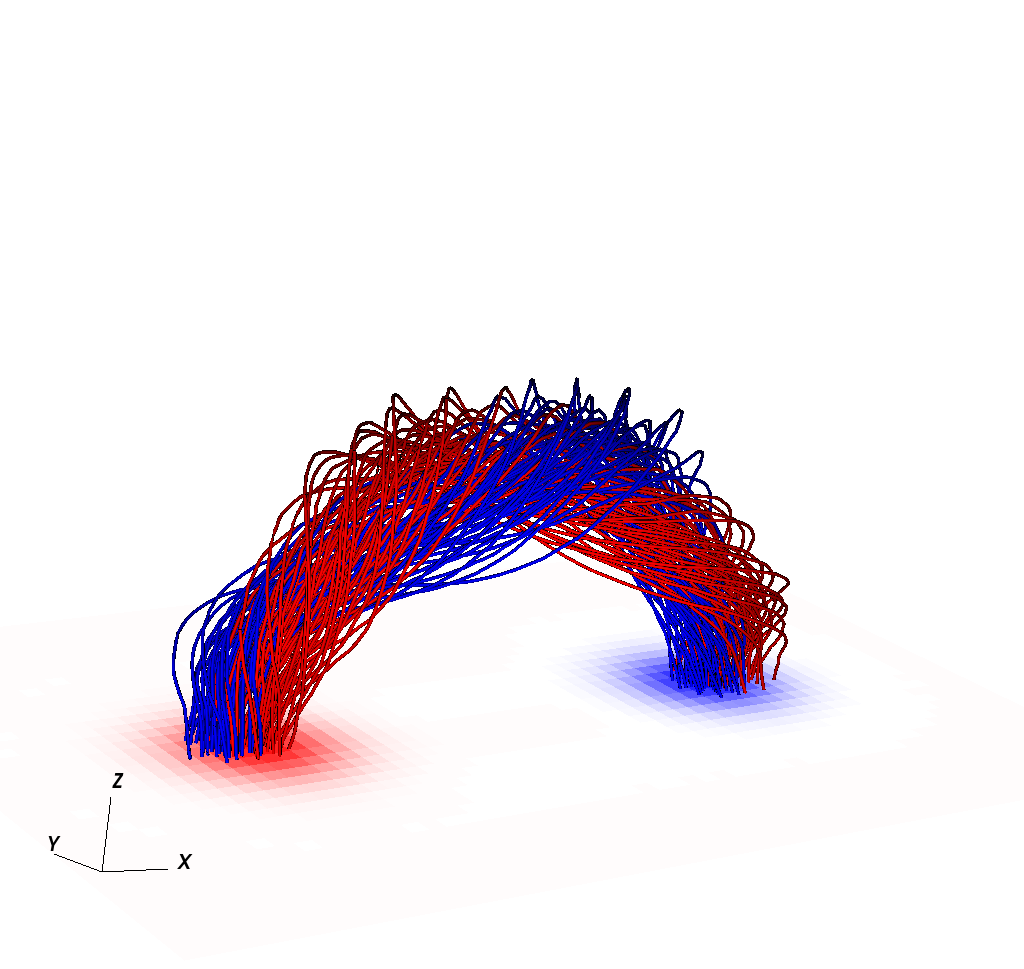}
		\caption{End of the rise time via $\mathbf{E}$-driving.}\label{fig:FG_rise_end_E_driving}
	\end{subfigure}
	
	\begin{subfigure}[b]{.45\linewidth}
		\includegraphics[width=\linewidth]{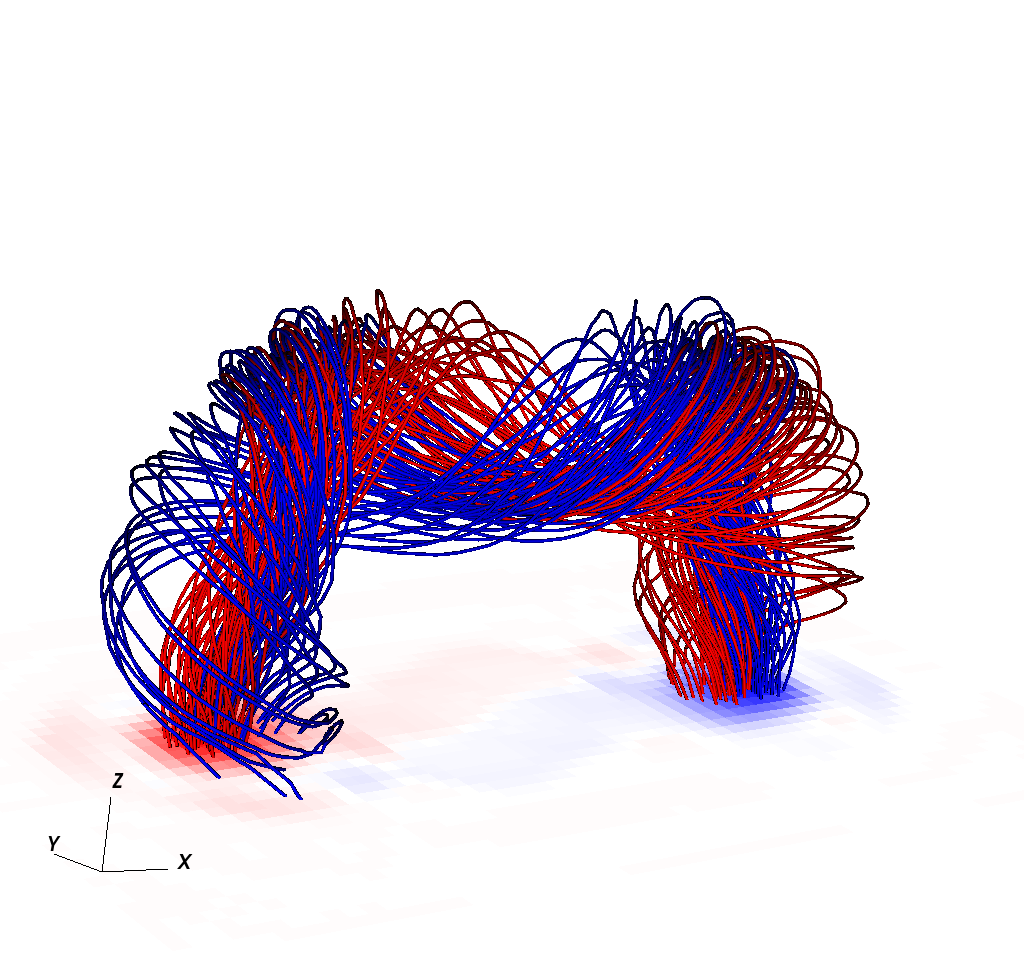}
		\caption{At t=108 time units (anchored for 58 time units) via $\mathbf{B}$-driving.}\label{fig:FG_long_B_driving}
	\end{subfigure}
		\begin{subfigure}[b]{.45\linewidth}
		\includegraphics[width=\linewidth]{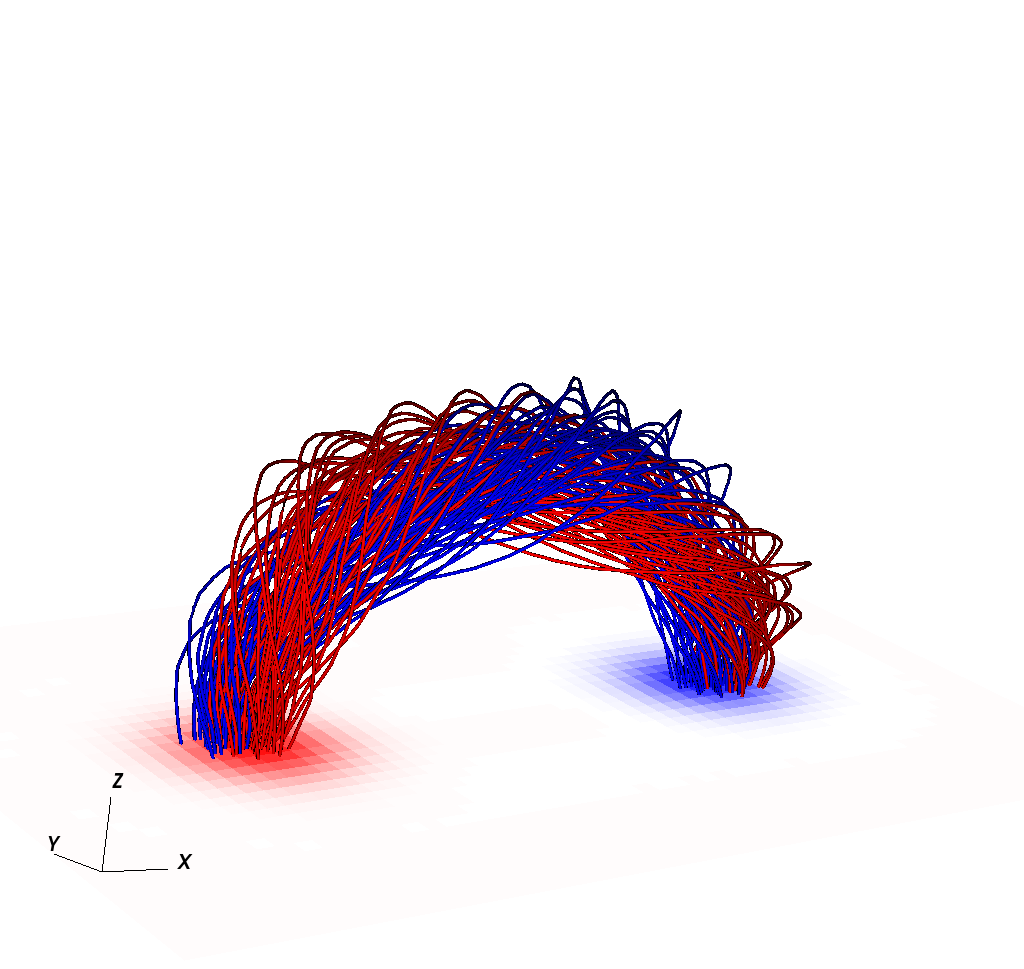}
		\caption{At t=108 time units (anchored for 58 time units) via $\mathbf{E}$-driving. }\label{fig:FG_long_E_driving}
	\end{subfigure}
	\caption{Comparison of the structure of the magnetic field lines in the flux tube at the end of the rise time due to, (a) $\mathbf{B}$-driving and (b) $\mathbf{E}$-driving, and at a later time after anchoring the footpoints for the Fan-Gibson flux emergence test case (see Sect.~\ref{sec:fangibson}). The red and blue field lines are traced from seed points centered in the positive (red) and negative (blue) polarities. The pseudocolour plot indicates $B_z$ .}
	\label{fig:fan_gibson_rise_end}
\end{figure*}
~\ref{fig:fan_gibson_rise_end} compares the flux-tube morphology at the end of the rise phase ($t=54$) and at a later time after the imposed rise has stopped, for the $\mathbf{B}$-driving and $\mathbf{E}$-driving simulations.
Comparison of Figures~\ref{fig:FG_long_B_driving} and \ref{fig:FG_long_E_driving} shows a dramatic example of the problems that can arise from an inappropriate data-driving strategy. 
In this case, $\mathbf{B}$-driving over-injects magnetic helicity and produces incorrect $B_z$ values of larger magnitude at the positive and negative polarities of the flux tube.
This error does not merely cause a quantitative departure from the expected flux-emergence solution. 
By over-injecting helicity, the $\mathbf{B}$-driven simulation also introduces excessive twist into the coronal magnetic field. 
The flux tube then becomes susceptible to the kink instability, which qualitatively changes the subsequent evolution. 
As a result, the $\mathbf{B}$-driven simulation no longer reproduces the intended test case: the magnetic structure evolves away from the analytical flux-emergence solution and enters a different dynamical regime.

Overall, both $\mathbf{B}$-driving and $\mathbf{E}$-driving perform well during the rise phase. 
After the rise stops, however, $\mathbf{E}$-driving remains much more accurate, maintaining both $H_R$ and $\Phi$ close to the analytical reference values. 
In contrast, $\mathbf{B}$-driving deviates significantly from the analytical result because the accumulated errors in helicity and boundary flux alter the magnetic configuration and trigger kink-unstable dynamics.
These errors are a direct manifestation of the fundamental limitations of the conventional ghost-cell $\mathbf{B}$-driving described in Section~\ref{sec:method_b_driving}---the supplied ghost-cell state is not calculated to reproduce the required boundary EMFs and leads to a systematic error/failure in this case. 
We emphasise that in this test case the driving data are complete---$\mathbf{B}$ and $\mathbf{v}$ are known analytically at every ghost-cell location and time---yet directly assigning these values still fails to reproduce the EMFs required for the desired boundary update. 
The deviation is therefore attributed to the boundary-condition implementation itself, not to data limitations. 
This issue is common in the literature because $\mathbf{B}$-driving is often implemented in a way that appeals to physical intuition. 



\section{Ground-truth Simulated Buoyantly Rising Flux-tube}
The simulation by \cite{toriumi2017numerical}---hereafter the ground-truth (GT) simulation---models the emergence of a buoyant flux-tube from the convection zone into the corona. 
Data (velocity and magnetic field vectors) extracted from a horizontal layer within this simulation corresponding to the photosphere are provided by \cite{toriumi2020comparative}.
Our interest here is in determining whether the data from this surface alone is sufficient to drive the same evolution above this plane when no other data is given. 
It has been shown by \cite{Leake_2017} that good agreement can be achieved if all physical variables are prescribed over all ghost-zone layers in the  $z$-dimension required to define the driving boundary, with sufficiently high cadence---the work reproduced a ground truth simulation using a data-driven variant of the same numerical solver. 
A similar analysis was completed by \citet{toriumi2020comparative}, which compared several different data-driven codes, reproducing the GT with varying degrees of success.
Our goals here are to explore (1) which subset of the GT boundary data and (2) what numerical implementation of driving yields plasma evolution that most accurately matches that of the GT.
The present study improves on the work of \citet{toriumi2020comparative} by comparing several variations of $\mathbf{B}$-driving and $\mathbf{E}$-driving strategies in otherwise identical conditions. 

The GT simulation gives a more realistic and computationally challenging test case than the example described in Section \ref{sec:fangibson}, but is still significantly better constrained than true observational data. 
With the GT simulation we have access to all three components of the magnetic field and the velocity. 
The current test case is therefore a good surrogate for testing which data driving strategy is numerically superior in the context of photospheric-like observations, although this does not include the uncertainty associated with inversions of $\mathbf{B}$, $\mathbf{E}$, or $\boldsymbol{v}$.
In this comparison the $\mathbf{B}$-driving is at a disadvantage because $\mathbf{B}$ and $\boldsymbol{v}$ are only known at a single $z$-height, i.e., the photosphere, but they are required at varying ghost-cell heights. 
Conversely, the $\mathbf{E}$-driving is at an advantage since only values at a single $z$-height ($E_x$ and $E_y$ at the $k=-1/2$ physical domain boundary) are required for driving the lower $z$-boundary.
Note that $\mathcal{E}_x$ and $\mathcal{E}_y$ directly constrain the $B_z$ evolution on the boundary ($k=-1/2$), but the horizontal magnetic field components are domain values ($k=1$) and are not guaranteed to match the driving data prescribed values on the boundary ($k=-1/2$).
The insights from this comparison are useful and important for the numerical methods used to simulate solar flux emergence considering that: (1) in our previous test case the knowledge of $\mathbf{B}$ and $\boldsymbol{v}$ at all spatial locations did not resolve differences between the driving strategies; (2) any driving strategy must contend with this issue through ad-hoc mitigating procedures \citep{Jiang2016,guo2019solar,Tarr2024MHDIII}; and (3) the observational reality only provides data at single $z$-height.

\subsection{Non-dimensionalisation}
A common issue faced when reproducing this GT simulation is calculating the correct background atmospheric profile based on the non-dimensionalisation reference values given in \cite{toriumi2017numerical}.
It is therefore appropriate to clarify this here. 
For better readability, we shall work in SI units throughout this subsection and the next.

The correct GT non-dimensionalisation is achieved by choosing the following primary reference parameters for the scale height $L_0$, density $\rho_0$, velocity $v_0$, and temperature $T_0$:
\begin{subequations}\label{eq:Toriumi_primary_refVals}
\begin{align}
    L_0 &= 170 \times 10^{3}\,\mathrm{m},  \\
    \rho_0 &= 1.4 \times 10^{-4}\,\mathrm{kg\,m^{-3}},  \\
    v_0 &= C_0 = \sqrt{\frac{p_0}{\rho_0}} = v_{\text{a},0} = \frac{B_0}{\sqrt{10 \rho_0}} = 6800\,\mathrm{m\,s^{-1}}, \label{eq:GT_v0} \\
    T_0 &= \frac{\bar{m} C_{0}^2}{\gamma k_\text{B}} = 5600\,\mathrm{K}, \label{eq:GT_T0}
\end{align}
\end{subequations}
where $C_0$, $p_0$, $v_{\text{a},0}$, and $B_0$ are, respectively, the reference isothermal sound speed, pressure, Alfv\'en speed, and magnetic field strength, $\bar{m}$ is the mean molecular mass, and $k_\text{B}$ is the Boltzmann constant. The $1/\sqrt{10}$ factor in Equation~(\ref{eq:GT_v0}) accounts for the conversion of the units used in \cite{toriumi2017numerical} into SI.
Any other reference parameter should now be treated as secondary and derived from these values.
Specifically, $B_0$, $p_0$, $\bar{m}$, and the reference gravitational acceleration $g_0$ evaluate to:
\begin{subequations}\label{eq:Toriumi_derived_refVals}
\begin{align}
    B_0 &= 0.025443~\mathrm{T},   \\
    p_0 &=  6473.6~\mathrm{Pa}, \\
    \bar{m} &= \frac{k_\text{B} T_0 \gamma}{C_0^2} = 2.78677 \times 10^{-27}\,\textrm{kg} \label{eq:GT_mbar}, \\
    g_0 &= C_0^2 / L_0 = 272 ~\mathrm{m}\,\mathrm{s}^{-2}.  \label{eq:GT_g}
\end{align}
\end{subequations}
This $B_0$ is the SI equivalent value used by \cite{toriumi2017numerical}, which follows from their definition of $v_{\text{A},0}$ given in Equation~(\ref{eq:GT_v0}). However, other definitions of $v_{\text{A},0}$ used in other codes (due to unit differences) can yield a different value, say, $B'_0$. Here the prime indicates a code-specific parameter different from the one used by \cite{toriumi2017numerical}. When expressed in the same physical units, this $B'_0$ can be different to the $B_0$ given above. For example, LareXd's internal nondimensionalization scheme demands that $B'_0 = v_{\text{A},0}\sqrt{\mu_0 \rho_0} = 0.0901$ T, where $\mu_0$ is the vacuum's magnetic permeability.
Thus, in order to get the correct nondimensional values, the input nondimensional GT magnetic field driving data must be rescaled by $B_0 / B'_0$.

\subsection{GT Background Hydro\-static Gas}\label{sec:appendix_atmosphere}
The background GT model is constructed based on the common piecewise temperature function given by Equations~10--12 of \cite{toriumi2017numerical}. 
In SI units, we have\footnote{Note that there is a typo in Equation~11 of \cite{toriumi2017numerical}, where the gravitational acceleration should be $g$ (not $g_0$).}:
\begin{equation}\label{eq:Toriumi_T_dim}
    T(z) =
\begin{cases}
T_\text{ph} - \frac{\gamma - 1}{\gamma}\frac{\bar{m} g}{k_\text{B}} z, & \text{if}\, z \leq 0,\\
T_\text{ph} + \frac{1}{2}\left(T_\text{cor} - T_\text{ph}\right) & \text{if}\, z \geq 0 \\
\qquad \times \left(\tanh\left(\frac{z - z_\text{cor}}{W_\text{tr}}\right) + 1\right), &
\end{cases}
\end{equation}
in which $T_\text{ph} = T_0$, $T_\text{cor} = 150 T_0$, $z_\text{cor} = 18 L_0$, and $W_\text{tr} = 2 L_0$.
The density and pressure profiles follow assuming the ideal gas equation of state
\begin{equation}\label{eq:Toriumi_EOS_dim}
    p = \frac{k_\text{B}}{\bar{m}} \rho T,
\end{equation}
and the hydrostatic model
\begin{equation}\label{eq:Toriumi_hydrostatic_dim}
    \dfrac{d p}{d z} + \rho g = 0.
\end{equation}
In what follows, we will demonstrate the correct nondimensionalization of these equations for Athena++ and LareXd. The same procedure may be applied to other codes.

Let us start by Equation~\ref{eq:Toriumi_T_dim}. In most codes (including Athena++ and LareXd) the internal nondimensionalization scheme demands that the temperature reference value is set by $T_0' = \bar{m} g_0 L_0 / k_\text{B}$. Substituting $g_0$ from Equation~\ref{eq:GT_g} yields $T_0' = \bar{m} C_0^2 / k_\text{B}$. Notice that this is different from Equation~\ref{eq:GT_T0} by a factor of $\gamma$, i.e., $T_0' = \gamma T_0$. 
Thus, as with the case of $B'_0$ explained above, the correct nondimensional temperature profile is achieved through $\hat{T}' = \left(T_0 / T_0'\right)\hat{T}$, where $\hat{\left(.\right)}$ denotes a dimensionless quantity.
To this end, we first non-dimensionalize Equation~\ref{eq:Toriumi_T_dim} using the reference value $T_0$ given by Equation~\ref{eq:GT_T0}, and then multiply the result ($\hat{T}$) by the reference re-scaling ratio $T_0 / T_0' = 1 / \gamma$. This yields the following dimensionless GT-consistent temperature model $\hat{T}'$ in our codes
\begin{equation}\label{eq:Toriumi_T_nondim}
    \hat{T}'(\hat{z}) = \frac{1}{\gamma}
\begin{cases}
1 - \left(\gamma - 1\right) \hat{g} \hat{z}, & \text{if}\, \hat{z} \leq 0,\\
1 + \frac{149}{2}\left(\tanh\left(\frac{\hat{z} - 18}{2}\right) + 1\right) & \text{if}\, \hat{z} \geq 0
\end{cases},
\end{equation}
wherein $\hat{g} = 1 / \gamma$ is the unitless gravitational acceleration. Importantly, note that this sets the physical gravitational acceleration used in the GT simulation to the reduced value of $g = g_0 \hat{g} = 163.2\,\mathrm{m.s}^{-2}$.

\begin{figure}
    \includegraphics[width=\linewidth]{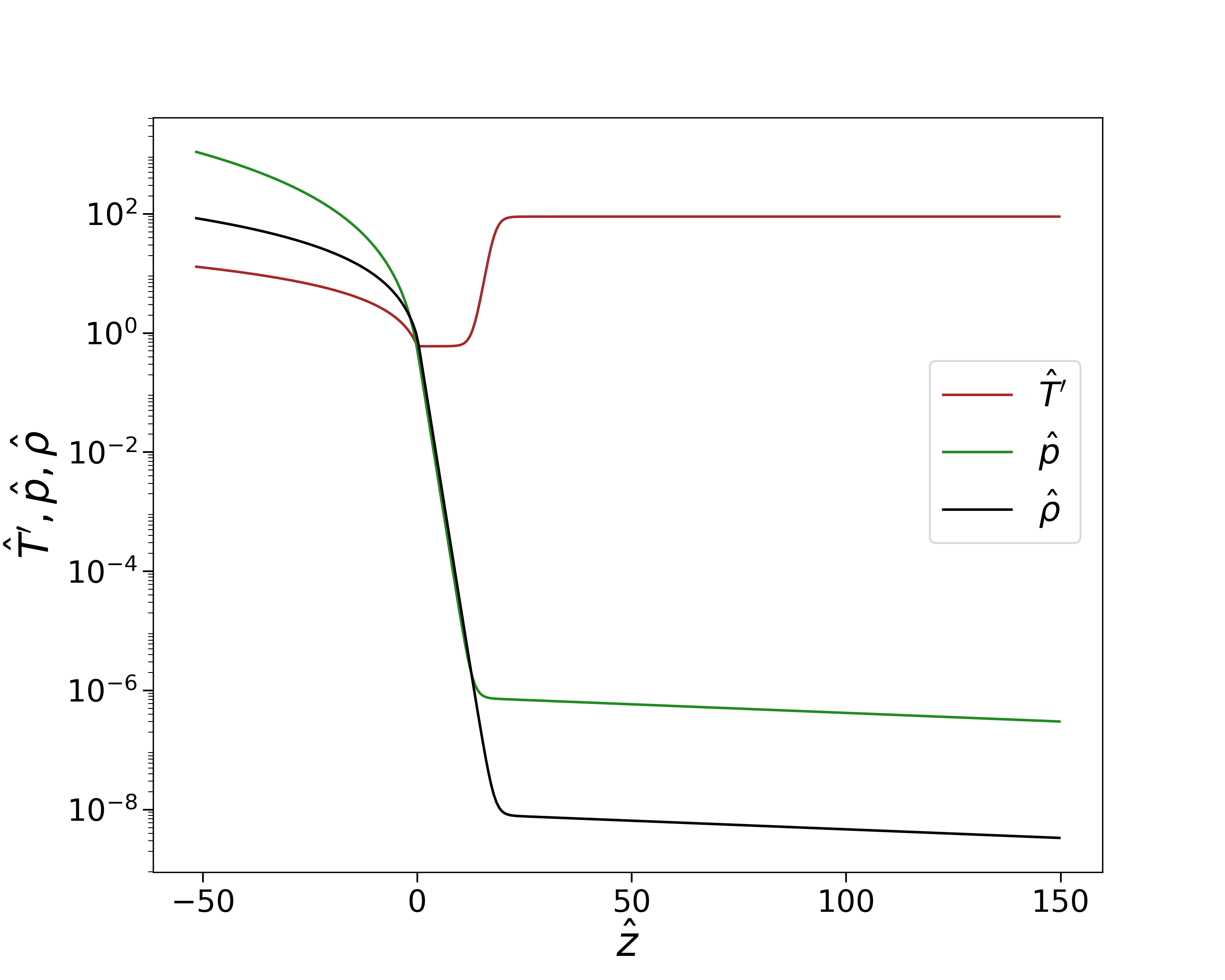}
    \caption{Initial thermodynamic profiles of our simulations in non-dimensional units. The gas model matches \cite{toriumi2017numerical}.}
    \label{fig:Toriumi_gas_nondim}
\end{figure}

Finally, using Equation~\ref{eq:Toriumi_T_nondim}, no further re-scaling is required for the density and pressure, as their reference values $\rho_0$ and $p_0$ are already in keeping with Athena++ and LareXd. Nondimensionalising Equations~\ref{eq:Toriumi_EOS_dim} and \ref{eq:Toriumi_hydrostatic_dim} using $T_0'$, $\rho_0$, and $p_0$, we find
\begin{equation}
    \hat{p} = \hat{\rho} \hat{T}'
\end{equation}
and
\begin{equation}\label{eq:GT_nondim_rho}
    \dfrac{d}{d\hat{z}}\left(\hat{\rho}\hat{T}'\right) + \hat{\rho} = 0,
\end{equation}
which are solved subject to the boundary condition $\hat{\rho}(\hat{z} = 0) = 1$. Figure~\ref{fig:Toriumi_gas_nondim} depicts the resulting gas model in nondimensional units, which matches Figure~2 of \cite{toriumi2017numerical}.

\subsection{Comparing Strategies and Results for the GT Simulation} \label{sec:data_driving_comparison_toriumi}
The driving strategies implemented for the comparison are summarised in Table 
\begin{table}
    \centering
    \begin{tabular}{|l|l|l|l|}
		\hline
        \hspace{-0.8cm}\begin{tabular}{l}Name \&\\ driving \end{tabular} & $\mathbf{B}$ update & \hspace{-0.8cm}\begin{tabular}{l}Velocity\\treatment \end{tabular}\\ \hline
        
		\hspace{-0.8cm}\begin{tabular}{l}E1\\$\mathbf{E}$-driving (DA) \end{tabular} & \hspace{-0.8cm}\begin{tabular}{l} Corrected EMFs, \\ $\mathcal{E}_x$ and $\mathcal{E}_y$ \end{tabular} &  \hspace{-0.8cm}\begin{tabular}{l} $\boldsymbol{v}$ data,\\linear recon.\end{tabular} \\ \hline
        
		\hspace{-0.8cm}\begin{tabular}{l}E2\\$\mathbf{E}$-driving (DA) \end{tabular}& \hspace{-0.8cm}\begin{tabular}{l} Corrected EMFs, \\ $\mathcal{E}_x$ and $\mathcal{E}_y$ \end{tabular}  & zero mass flux \\ \hline
        
		\hspace{-0.8cm}\begin{tabular}{l}E3\\$\mathbf{E}$-driving (DA) \end{tabular} & \hspace{-0.8cm}\begin{tabular}{l} Corrected EMFs, \\ $\mathcal{E}_x$ and $\mathcal{E}_y$ \end{tabular}  & zero-gradient \\ \hline
        
		\hspace{-0.8cm}\begin{tabular}{l}B1\\$\mathbf{B}$-driving (DD) \end{tabular} & \hspace{-0.8cm}\begin{tabular}{l}$\mathbf{B}$ data \\ constant in z\end{tabular} & \hspace{-0.8cm}\begin{tabular}{l} $\boldsymbol{v}$ data,\\linear recon.\end{tabular} \\ \hline
		\hspace{-0.8cm}\begin{tabular}{l}B2\\$\mathbf{B}$-driving (DD) \end{tabular}& 
        \hspace{-0.8cm}\begin{tabular}{l}$\nabla\cdot\mathbf{B}=0$ enforced\\ in ghost zones \end{tabular}
        & \hspace{-0.8cm}\begin{tabular}{l} $\boldsymbol{v}$ data,\\linear recon.\end{tabular} \\ \hline
	\end{tabular}
    \caption{Data driving strategies used in the reproduction of the buoyantly rising flux-tube ground truth simulation by \citet{toriumi2020comparative}.
    DD and DA indicate data driving and data assimilation techniques, see Section~\ref{sec:frame_data_driving}. 
    }
\label{table:toriumi_driving_variations}
\end{table}
\ref{table:toriumi_driving_variations}.
All E-driving strategies prescribe the same corrected boundary EMFs, obtained from the superposition $\boldsymbol{\mathcal{E}}=\boldsymbol{\mathcal{E}}^{C}+\boldsymbol{\mathcal{E}}^{vB}$ (as described in Section \ref{sec:method_e_driving}). 
The primary electric field is calculated from the GT data via an ideal Ohm's Law. 
E1, E2, and E3 are therefore a data-assimilation strategy in the terminology of Section~\ref{sec:driving_vs_assimilation}, whereas B1 and B2 are pure data driving. 
The different $\mathbf{B}$-driving strategies considered are (1) copying the 2D driving data plane for $\mathbf{B}$ into each plane of ghost-cells (labelled ``$\mathbf{B}$ constant $z$''), 
or (2) a modified $\mathbf{B}$ to enforce $\nabla\cdot\mathbf{B}$ by manipulating $B_z$ on the outer ghost-zone cell-face.
Two other $\mathbf{B}$-driving strategies were investigated to complement the E2 and E3 strategies but are not discussed here due to physically incorrect evolutions (see Appendix~\ref{sec:Bdriving_innapropriate} and Figure~\ref{fig:B_driving_velocity_comparison_phi} for details). 
The velocity boundary conditions in the driving strategies are either (1) a linear extrapolation of velocity into the ghost-zones (using domain and GT driving data), (2) zero mass-flux, or (3) zero-gradient. 

Relative magnetic helicity and other metrics are calculated as described in Section~\ref{sec:data_driving_comparison_fan} and are shown in Figure
\begin{figure*}
    \plotone{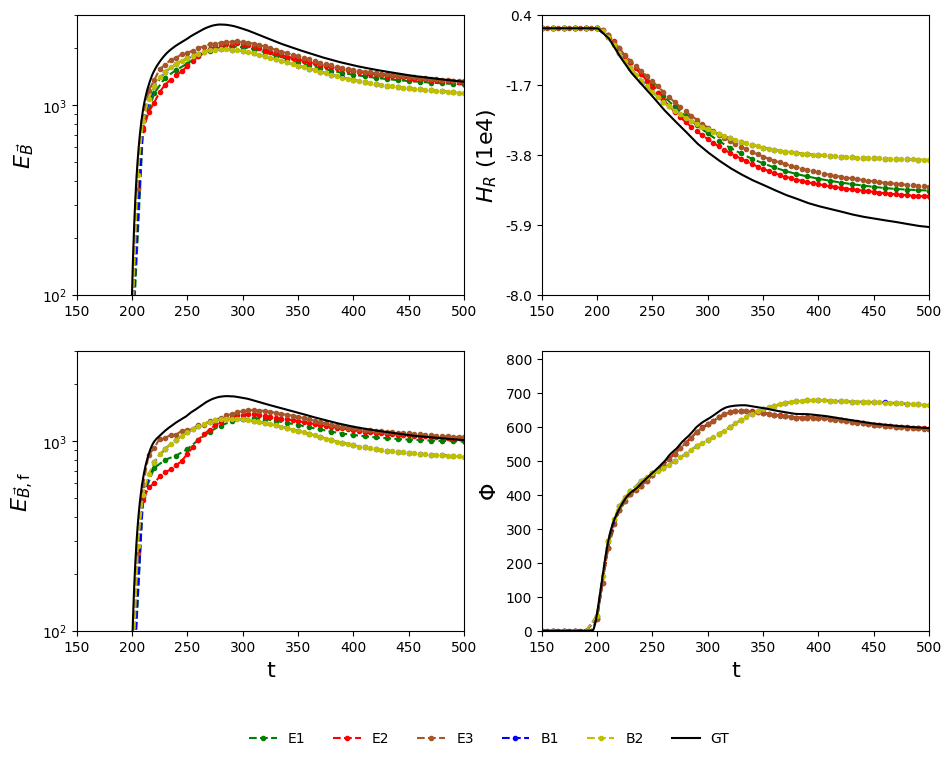}
	\caption{Magnetic energy (top left), magnetic free energy (bottom left), relative magnetic helicity (top right), and total unsigned flux on the driven boundary (bottom right). See Table~\ref{table:toriumi_driving_variations} for details: E1, E2, and E3 are E-driven and differ only in their velocity boundary conditions (data, zero mass flux, and zero-gradient, respectively). B1 and B2 are conventional $\mathbf{B}$-driving strategies: B1 assigns the magnetic-field data uniformly across the ghost cells, B2 adjusts the ghost-cell field to enforce the discrete solenoidal constraint. The ground truth results (GT, black) are the reference/target solution. 
    \label{fig:driving_comparison_Em_Emf_Hr_Phi}}
\end{figure*} 
\ref{fig:driving_comparison_Em_Emf_Hr_Phi} and
\begin{figure}
    \plotone{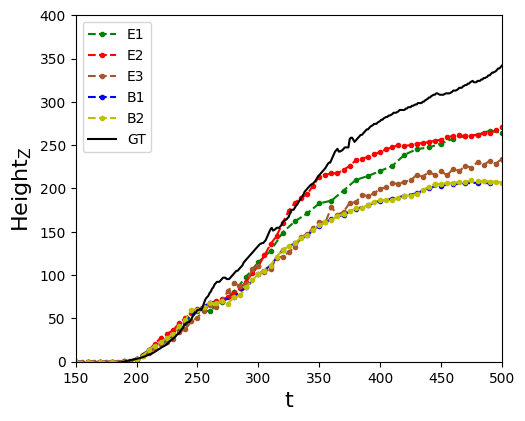}
	\caption{Height of flux-tube from data-driven simulations using the ground-truth data set \citep{toriumi2020comparative} with different Athena++ driving strategies. See Table~\ref{table:toriumi_driving_variations} for descriptions of symbols in the legend. The ground truth (GT) results (black) are the reference/target solution.  \label{fig:driving_comparison_height}}
\end{figure} 
\ref{fig:driving_comparison_height}. 
%
\begin{figure*}
	\centering
	\begin{subfigure}[b]{.3\linewidth}
		\includegraphics[width=\linewidth]{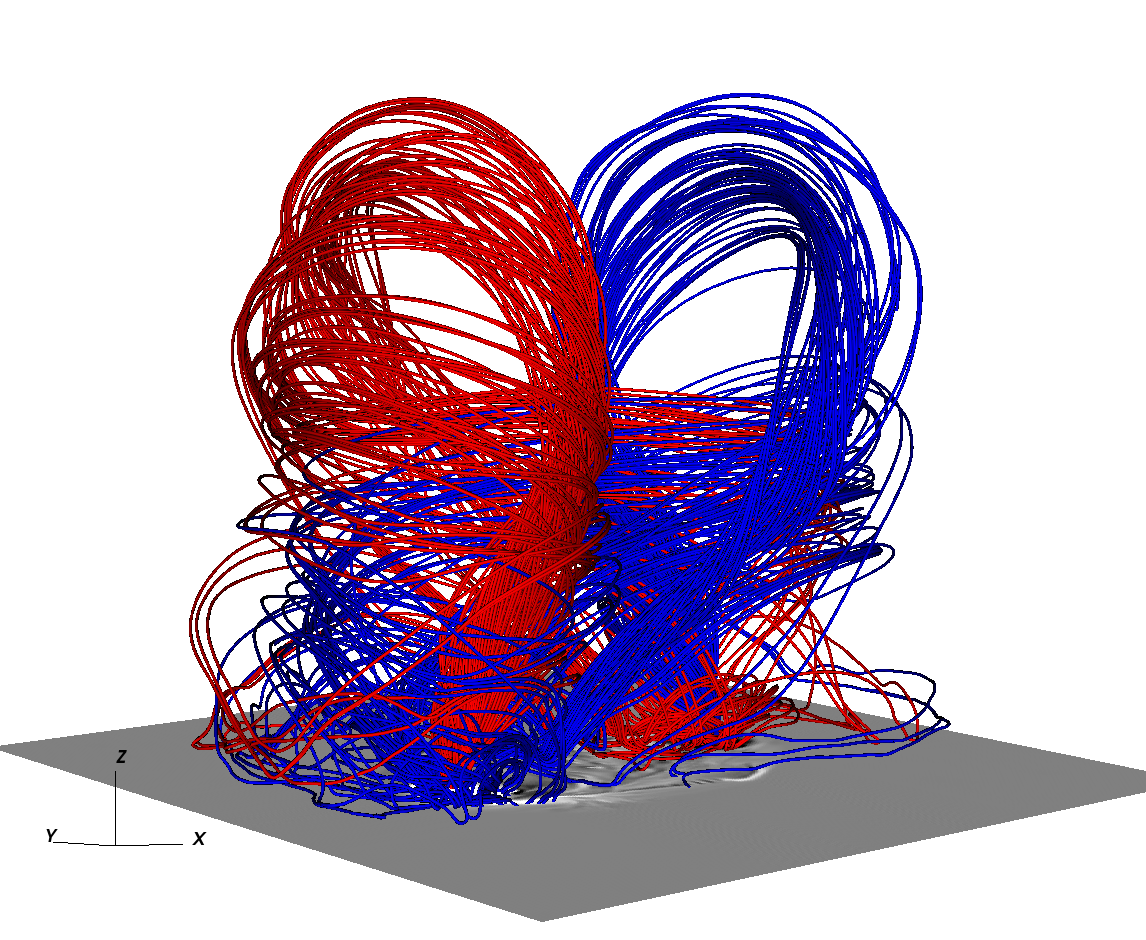}
		\caption{E1}\label{fig:GT_E1_lines}
	\end{subfigure}
	\begin{subfigure}[b]{.3\linewidth}
		\includegraphics[width=\linewidth]{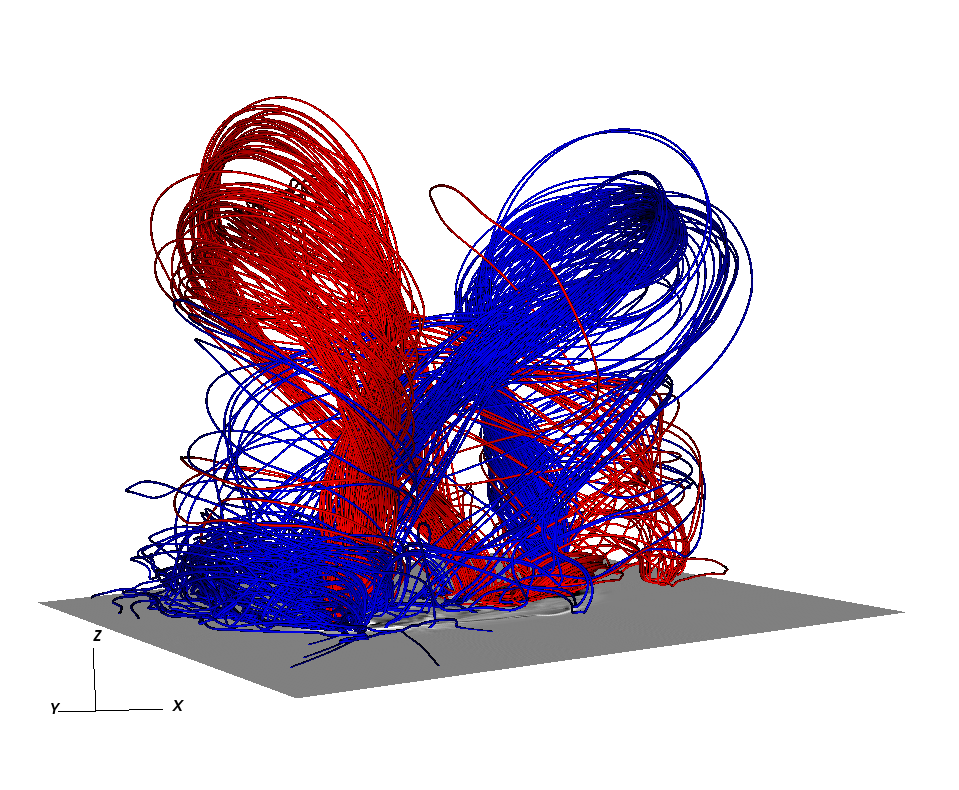}
		\caption{E2}\label{fig:GT_E2_lines}
	\end{subfigure}
	\begin{subfigure}[b]{.3\linewidth}
		\includegraphics[width=\linewidth]{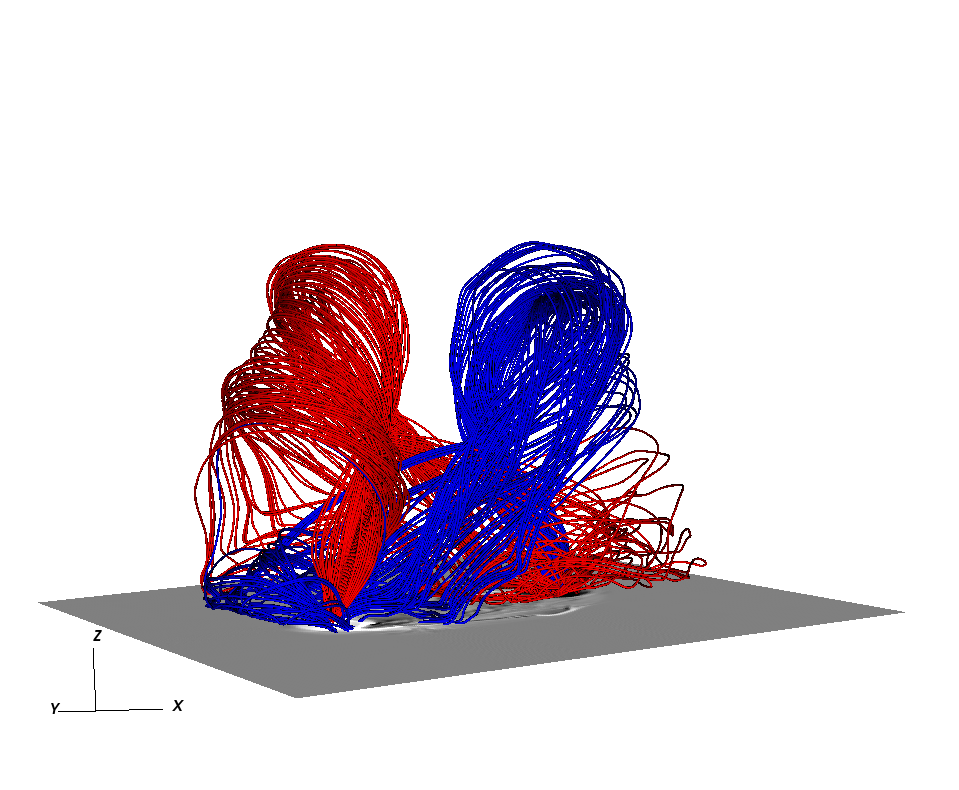}
		\caption{E3}\label{fig:GT_E3_lines}
	\end{subfigure}
    
    \begin{subfigure}[b]{.3\linewidth}
		\includegraphics[width=\linewidth]{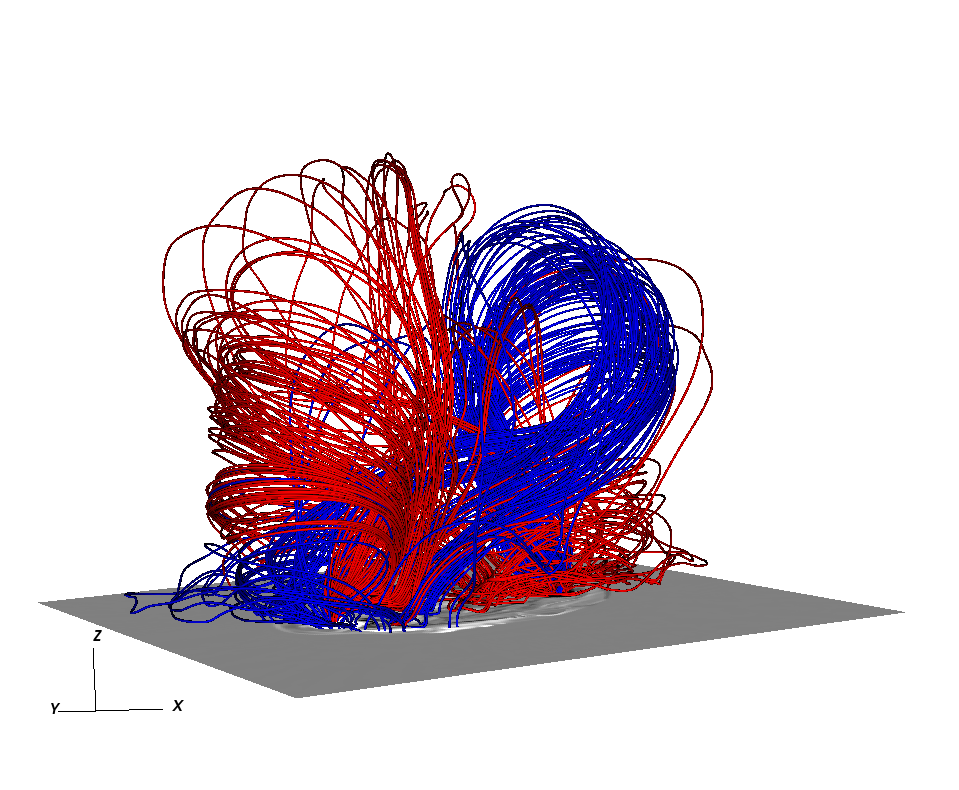}
		\caption{B1}\label{fig:GT_B1_lines}
	\end{subfigure}
    \begin{subfigure}[b]{.3\linewidth}
		\includegraphics[width=\linewidth]{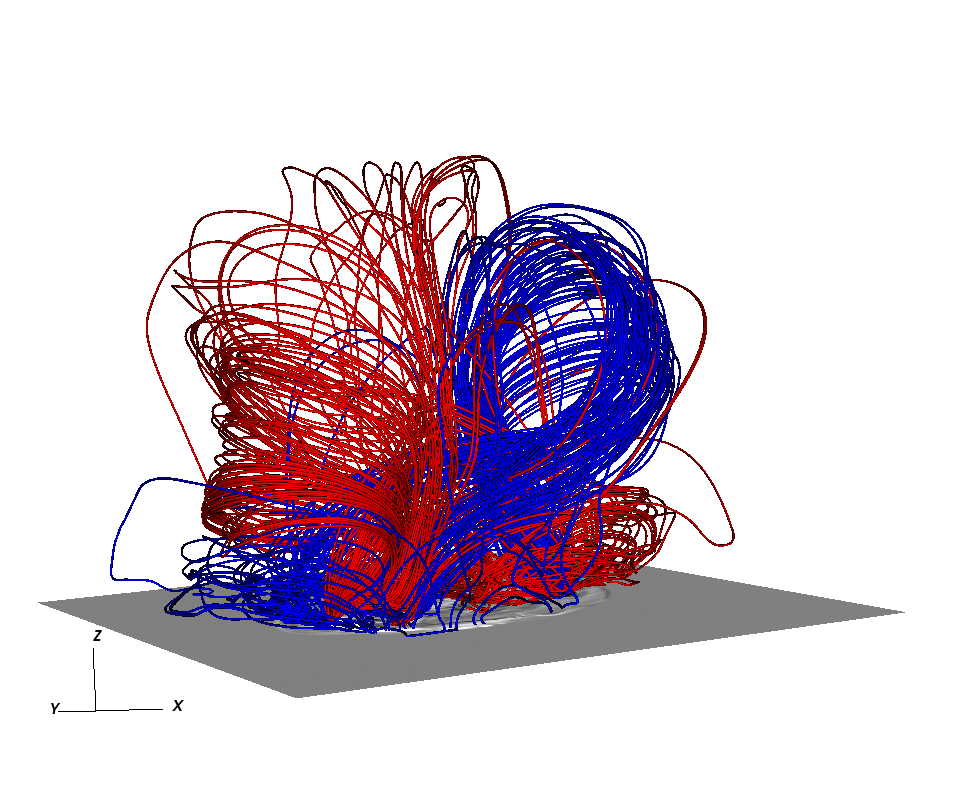}
		\caption{B2}\label{fig:GT_B2_lines}
	\end{subfigure}
	\caption{Isometric view of magnetic field line traces for each data-driving strategy's reproduction of the ground-truth (GT) simulation. See Table~\ref{table:toriumi_driving_variations} for details: E1, E2, and E3 are E-driven and differ only in their velocity boundary conditions (data, zero mass flux, and zero-gradient, respectively). B1 and B2 are conventional $\mathbf{B}$-driving strategies: B1 assigns the magnetic-field data uniformly across the ghost cells, B2 adjusts the ghost-cell field to enforce the discrete solenoidal constraint. The ground truth results (GT, black) are the reference/target solution. \label{fig:driving_comparison_field_lines_iso}}
\end{figure*}
\begin{figure*}[ht!]
	\centering
	\begin{subfigure}[b]{.3\linewidth}
		\includegraphics[width=0.875\linewidth]{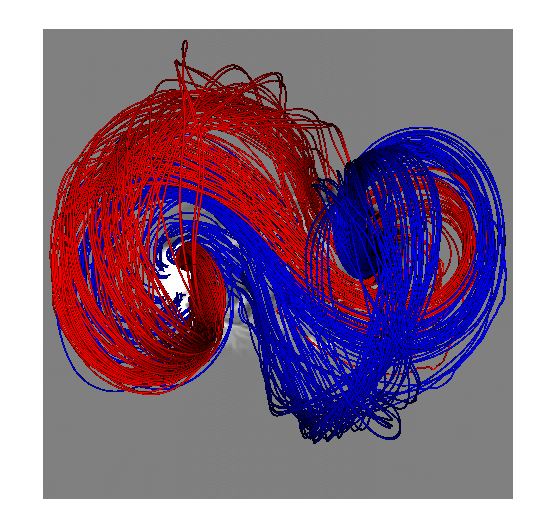}
		\caption{E1}\label{fig:GT_E1_lines_top}
	\end{subfigure}
	\begin{subfigure}[b]{.3\linewidth}
		\includegraphics[width=\linewidth]{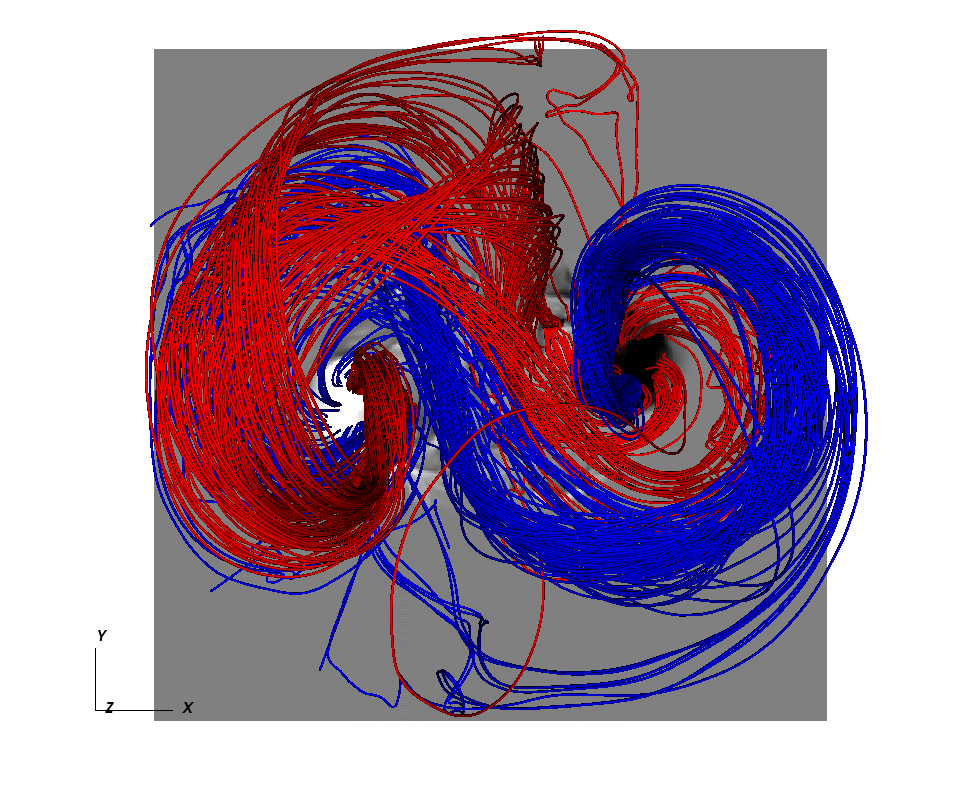}
		\caption{E2}\label{fig:GT_E2_lines_top}
	\end{subfigure}
	\begin{subfigure}[b]{.3\linewidth}
		\includegraphics[width=\linewidth]{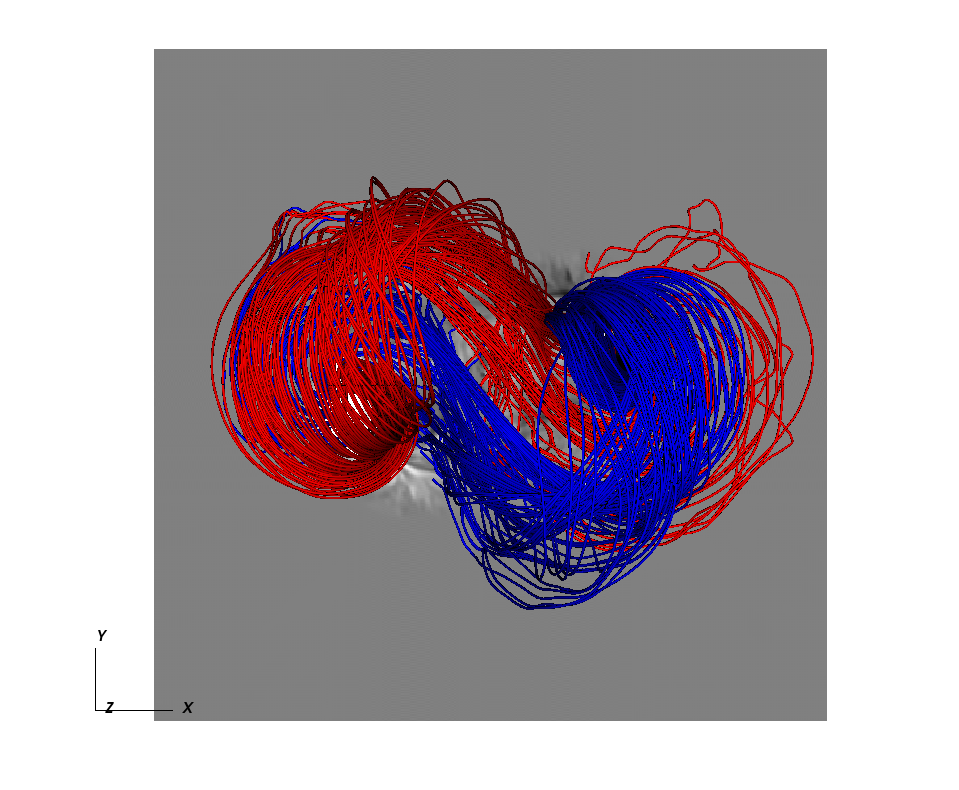}
		\caption{E3}\label{fig:GT_E3_lines_top}
	\end{subfigure}

    \begin{subfigure}[b]{.3\linewidth}
		\includegraphics[width=\linewidth]{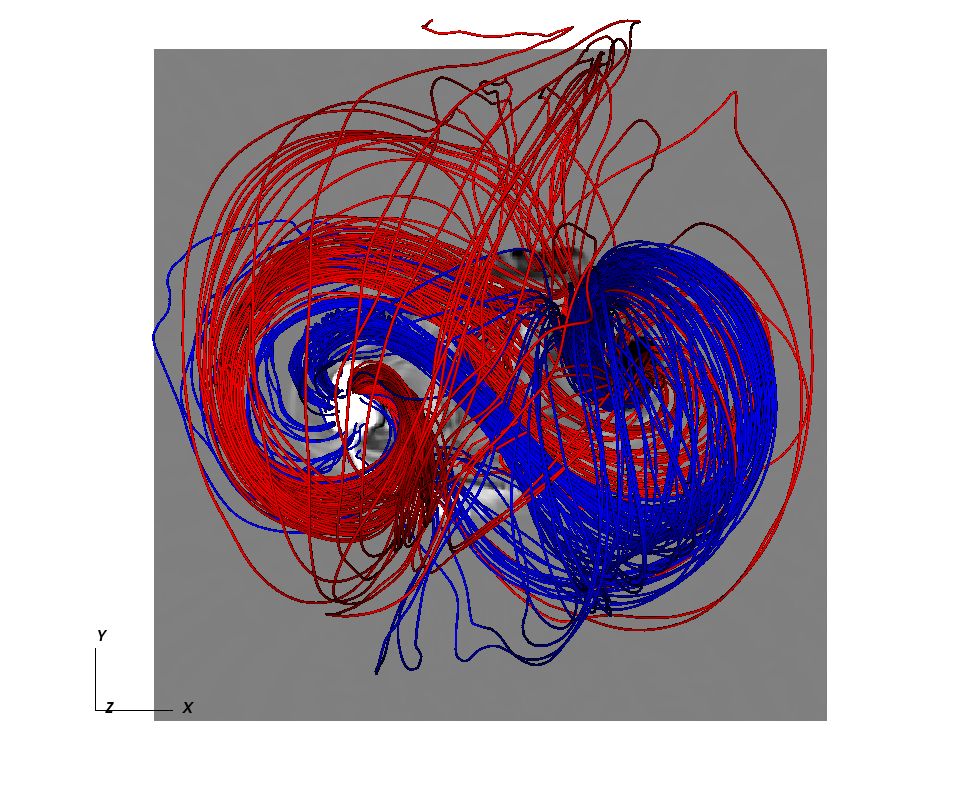}
		\caption{B1}\label{fig:GT_B1_lines_top}
	\end{subfigure}
    \begin{subfigure}[b]{.3\linewidth}
		\includegraphics[width=\linewidth]{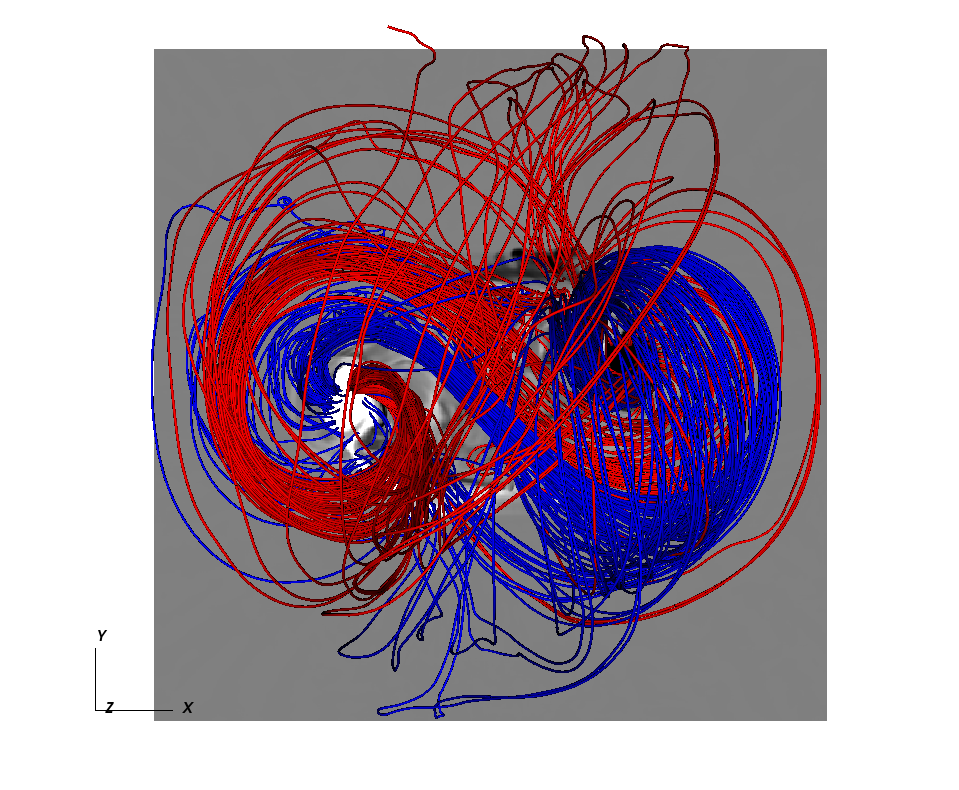}
		\caption{B2}\label{fig:GT_B2_lines_top}
	\end{subfigure}
	\caption{Top-down view of magnetic field line traces for each data-driving strategy's reproduction of the ground-truth (GT) simulation. See Table~\ref{table:toriumi_driving_variations} for details: E1, E2, and E3 are E-driven and differ only in their velocity boundary conditions (data, zero mass flux, and zero-gradient, respectively). B1 and B2 are conventional $\mathbf{B}$-driving strategies: B1 assigns the magnetic-field data uniformly across the ghost cells, B2 adjusts the ghost-cell field to enforce the discrete solenoidal constraint. The ground truth results (GT, black) are the reference/target solution. \label{fig:driving_comparison_field_lines_top}}
\end{figure*}
The figures show a time series of the relevant metrics for each driving strategy and the GT results from \citet{toriumi2020comparative}) as a reference solution.
Figures \ref{fig:driving_comparison_field_lines_iso} and \ref{fig:driving_comparison_field_lines_top}
show traces of magnetic field lines in the isometric and top-down perspectives, respectively.

In general, the $\mathbf{E}$-driven strategies E1 and E2 perform best in the sense that the quantities in Figures~\ref{fig:driving_comparison_Em_Emf_Hr_Phi} and~\ref{fig:driving_comparison_height} 
are closest to the GT time series. 
In terms of the morphology of the emerged flux-tube, E1 and E2 best reproduce the GT structures (see Figures 4 and 5 of \citet{toriumi2020comparative}) and the GT flux-tube height (Figure \ref{fig:driving_comparison_height}). 

Regarding $H_R$ and $\Phi$ in 
Figure~\ref{fig:driving_comparison_Em_Emf_Hr_Phi}
E1, E2, and E3 closely reproduce the total unsigned flux and they perform better in reproducing the relative magnetic helicity, $H_r$, than the B-driven results---though all series deviate. 
The GT $H_r$ values are best reproduced by E2, with similar but slightly less accurate results by E1 and E3. 
The B1 and B2 results deviate significantly from the GT $H_r$ series but all results follow the GT trend. 

Regarding $E_{\mathbf{B}}$ and $E_{\mathbf{B},\textrm{f}}$ in
Figure~\ref{fig:driving_comparison_Em_Emf_Hr_Phi}:
the magnetic energy profiles for all simulations follow the GT trend reasonably well, though with a reduced peak energy at an earlier time. 
The E3 result follows the GT magnetic energy best, which is interesting given that the only driving input is the $\mathbf{E}$-field values, whereas velocity is given by a zero-gradient extrapolation from within the domain---this suggests the dynamics of the magnetic field are adversely affected by the velocity driving at early time. 
Results across other strategies are more variable for magnetic free energy: the E1 and E2 results depart from the GT time series at early times ($t\leq275$), with a reduction in free energy that then returns to a value closer to the GT value at later times. 
The E3 result again follows the GT trend more closely, also matching GT values at late times. 
The B1 and B2 results follow the GT series with a significant offset during the entire simulation.

The deviation in energy injection between $E$-driven data-driving results is likely due to the thermodynamic boundary conditions. 
The thermodynamic boundary conditions (i.e. the treatment of density, velocity, and pressure) are identical except for velocities. 
However, the fluxes depend on all thermodynamic properties being consistent with each other which is not guaranteed by enforcing the velocity driving data (from the GT simulation) with the Dirichlet boundary condition on density and pressure that is constant in $x$ and $y$. 
Improving the data-driving treatment of the density and pressure is an important objective of our future work. 
Inaccurate magnetic energy injection at early times will likely be detrimental to accurate simulation of solar eruptions when attempting to reproduce observations. 
To mitigate this energy discrepancy we have explored how energy injection can be manipulated in Section~\ref{sec:adhoc_energy_athena}. 

\subsection{Discussion of Driving Strategies}
In reproducing the GT simulation of \citet{toriumi2017numerical} in Athena++, $\mathbf{E}$-driving provides more accurate results than $\mathbf{B}$-driving. 
This is fundamentally due to the B-driving strategies implemented here (in Athena++) applying ``approximate'' data-driving boundary conditions for the magnetic field evolution, as described in Section~\ref{sec:method_b_driving}. 
As established in Section~\ref{sec:method_b_driving}, the update cannot be controlled exactly by specifying ghost-cell values of $\mathbf{B}$ and $\mathbf{v}$ from a single data slice; in an Eulerian CT algorithm the boundary EMFs govern the update exactly, and E-driving specifies these directly.

Of the $\mathbf{E}$-driving strategies, E1 and E2 produce good overall agreement with GT results but we emphasise one obvious limitation of the latter: E2 does not allow for mass-flux at the lower boundary which is inconsistent with the GT simulation. 
E2 performs well in our comparisons because our assessment metrics are focused on the magnetic field evolution. 
The applications targeted by our driving strategies, namely solar eruptions and their space-weather impacts, require accurate treatment of both plasma thermodynamics and magnetic-field evolution. 
Therefore, E2 should be avoided when suitable alternatives are available, so we endorse E1 as the preferred driving strategy out of those investigated here.




Overall, these results indicate (1) $\mathbf{E}$-driving is preferable to conventional $\mathbf{B}$-driving in these particular circumstances, and (2) the treatment of thermodynamic variables in these data-driving strategies may limit their accuracy. 
Point (2) is supported by the varied performance of the $\mathbf{E}$-driving strategies' energy metrics (given only varying velocity boundary conditions), and the constant density and pressure inflow boundary conditions used. 
The expansion of the magnetic field into the surrounding atmosphere is affected by the initial background atmosphere, domain extents, and flux-tube thermodynamic properties. 
Across all scenarios simulated, the first two influences are constant and the same boundary EMFs are prescribed, and hence the same numerical fluxes of the magnetic-field components are imposed at the lower boundary.
Therefore, the fluxes of thermodynamic properties on the lower boundary are the most likely remaining cause of the variation in energy metrics at early time. 
These thermodynamic fluxes are influenced by density and pressure boundary conditions, but these are constant in this study---the varying velocity boundary condition will still cause the thermodynamic fluxes to vary across cases. 
Future work should investigate more sophisticated treatment of the boundary conditions for density and pressure; specifically it should implement a condition that accounts for the variation of density and pressure over the boundary with respect to flux-tube and surrounding inflow thermodynamic properties.

\section{Data-driving in Alternative Algorithms}\label{sec:cross_code_comparison}

In this section, we compare the best performing data driving strategy in Athena++ (E1) to several other strategies
(Figure~\ref{fig:driving_comparison_2020_paper_Em_Emf_hr_phi}). 
\citet{toriumi2020comparative} compared a number of data-driving strategies that were each implemented in different simulation tools, out of which we present simulation results from a magneto\-frictional (MF) \citep{cheung2012method} code and an MHD code \citep{hayashi2018mhd}, both using $\mathbf{E}$-driving. 
We also include novel results achieved from our implementation of $\mathbf{B}$-driving in LareXd \citep{ARBER2001151}, as well as an E1 Athena++ simulation with an arbitrary energy and helicity injection scheme.
Subsection~\ref{sec:data_drivinmg_numerical_mhd_errors} discusses numerical and physical consistency errors in data driving (in practice these appear tolerable from the reproduction of GT results) that researchers should be aware of.

\subsection{$\mathbf{E}$-driven GT simulations from Toriumi et al. (2017)}%

The MF implementation \citep{cheung2012method} (row one in Table~\ref{table:driving_methods_literature}) models the quasi-static evolution of magnetic fields---this acts as ``best case'' in reproducing relative magnetic helicity and total unsigned flux on the boundary. 
$H_R$ and $\Phi$ in Figure~\ref{fig:driving_comparison_2020_paper_Em_Emf_hr_phi}
show the excellent agreement of the MF results with GT simulation. 
As a result of neglecting inertial, gravity, and pressure effects, the MF model performs poorly when reproducing properties that interact with any thermodynamic processes, i.e. system energies ($E_{\mathbf{B}}$ and $E_{\mathbf{B},\textrm{f}}$ in 
Figure~\ref{fig:driving_comparison_2020_paper_Em_Emf_hr_phi})
and flux-tube expansion against the surrounding atmosphere.
The magnetic field update evolves the vector potential directly, with the boundary driven using $\mathbf{E}$-driving---the former makes the magnetic field divergence-free and the latter gives the advantages discussed in this work.

The $\mathbf{E}$-driven MHD model \citep{hayashi2018mhd} (row eight in Table~\ref{table:driving_methods_literature}) uses a different $\mathbf{E}$-driving than the implementation here for Athena++. 
The simulation over-injects helicity and energy significantly (see Figure~\ref{fig:driving_comparison_2020_paper_Em_Emf_hr_phi}).
Additionally, the morphology of the magnetic field deviates from the GT result, see Figure 5 of \citet{toriumi2020comparative}. 
This serves to illustrate that electric field driving is not necessarily an optimal method, depending on the implementation. 
The method of \citet{hayashi2018mhd} uses a sum of three electric field contributions computed at different locations within the boundary cells, and is
designed to reproduce the full observed evolution of $\mathbf{B}$ while maintaining $\nabla\cdot\mathbf{B}=0$. 
The \citet{hayashi2018mhd} simulation does not include a stratified atmosphere or gravity, which will cause deviations in thermodynamic properties like flux-tube height (influenced by the ability of the flux tube to expand against the surrounding atmosphere) but it does not account for deviations in $H_R$ and $\Phi$. 
Since the $\mathbf{E}$-driving dictates the evolution of the magnetic field on the lower boundary, the over-injection of magnetic energy and helicity, as well as deviation in the total unsigned flux on the boundary, indicate the data driving strategy is responsible for a significant amount of the deviation from the GT results.

\subsection{LareXd \textbf{B}-driven GT simulation: Implementation and Results}\label{sec:larexd_implementation_results}
The constrained transport algorithm used in LareXd (applied during the Lagrangian remap to the Eulerian grid) ensures divergence errors to machine precision.
However, the Lagrangian form of the induction equation solved in LareXd makes implementing an $\mathbf{E}$-driving strategy algorithmically challenging. Instead, following \cite{Leake_2017}, our implementation of data driving in LareXd uses $\mathbf{B}$-driving. 
This is done by prescribing the ghost-cell values at \texttt{zc(-1:0)} and \texttt{zb(-2:-1)}, as well as the boundary plane at \texttt{zb(0)}, where \texttt{zc} and \texttt{zb} (in LareXd notation) denote respectively the cell-centre and cell-boundary grid coordinates along the $z$-axis. Note that same as \cite{Leake_2017},  in the version of LareXd used in this study \texttt{zb(0)} is the first domain plane, and hence it is evolved.
However, unlike \cite{Leake_2017} who had access to the full 3D driving data slabs, our GT driving data are strictly 2D in spatial dimensions. 
Thus, we follow the same approach as outlined in Section~\ref{sec:method_b_driving}, and simply copy the driving data---after appropriate interpolations onto the LareXd staggered grid---into the above-mentioned boundary planes. 
We note that this prescription of the data-driver, while simple to implement, comes at the cost of accumulating field divergence errors at the bottom boundary. 
Nevertheless, the evolved field in the simulation domain above this plane remains divergence-free. 
Future work will aim to develop self-consistent data driving strategies better suited for Lagrangian remap codes like LareXd, i.e., without enforcing values on the boundary plane at \texttt{zb(0)}.

\begin{figure*}
\plotone{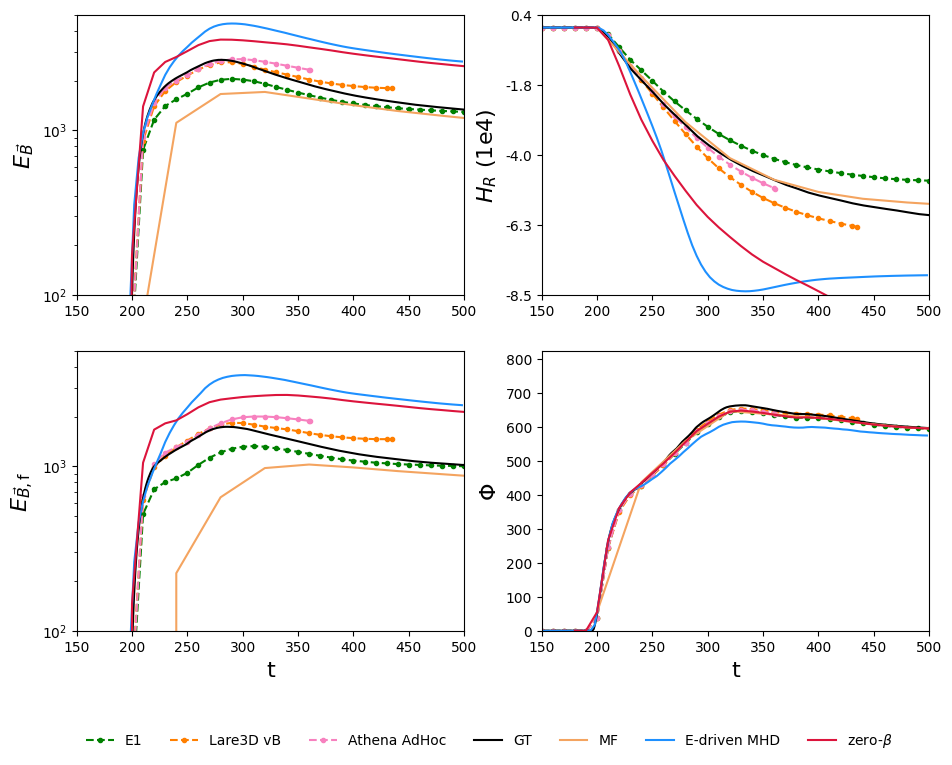}
	\caption{Magnetic energy (top left), magnetic free energy (bottom left), relative magnetic helicity (top right), and total unsigned flux on the driven boundary (bottom right). The E-1 series (labelled ``Athena vE'') is compared with the E-driven MF, $\mathbf{E}$-driven MHD, and ground-truth (GT) result \citep{toriumi2020comparative} and our $\mathbf{B}$-driven LareXd results. An $\mathbf{E}$-driven Athena simulation with ad-hoc energy injection is also compared for early time.     \label{fig:driving_comparison_2020_paper_Em_Emf_hr_phi}}
\end{figure*}

The LareXd results are shown as dashed orange curves with filled circles in 
Figure~\ref{fig:driving_comparison_2020_paper_Em_Emf_hr_phi}.
As seen in the $H_R$ panel of 
Figure~\ref{fig:driving_comparison_2020_paper_Em_Emf_hr_phi},
the driver seems to over-inject relative helicity---showing close initial agreement with the GT up to $t \approx 270$, but deviating thereafter. 
The high-precision agreement in the total unsigned flux through the boundary plane $\Phi(z = \texttt{zb(0)})$ seen in the $\Phi$ panel of 
Figure~\ref{fig:driving_comparison_2020_paper_Em_Emf_hr_phi}
is the direct consequence of the prescription of the ($B_z$) driving data at \texttt{zb(0)}, which may seem trivial. 
However, we have verified that $\Phi(z > \texttt{zb(0)})$ (not shown) closely follows a similar trend (as $\Phi(\texttt{zb(0)})$), with its amplitude diminishing with $z$ as expected.
 For $E_{\vec{B}}$ and $E_{\vec{B},\text{free}}$ depicted in 
 Figure~\ref{fig:driving_comparison_2020_paper_Em_Emf_hr_phi}
 the LareXd results show very good early time agreement ($t\lesssim 300$) before significantly deviating from the GT curves at late time. This is likely correlated with the over-injection of helicity. 
Possible causes may include the differing simulation box sizes, as well as code-specific discrepancies between the implementation of certain physical effects. 
For example, for the LareXd runs we have used a current-dependent resistivity profile, and it is worth noting that later phases of the simulation will inevitably involve magnetic reconnection \citep[][]{pontin2022}, whose onset and evolution is dependent on the nature of the (numerical) dissipation used \citep[][]{rembiasz2017}. 

Overall, the LareXd $\mathbf{B}$-driving method performs quite well despite divergence errors at the driving boundary. 
In comparison with E1, it performs slightly better in reproducing $H_\mathrm{R}$: while E1 undershoots $H_\mathrm{R}$ with early deviations from $t \approx 230$, LareXd overshoots $H_\mathrm{R}$ from $t \approx 270$ onwards. 
In terms of $\Phi$, both E1 and LareXd reproduce the correct results, albeit E1 has the edge over the $\mathbf{B}$-driven LareXd results since it self-consistently constrains $B_z$ at the bottom boundary. 
The E1 results show better reproduction of the GT flux tube height than LareXd, but both deviate significantly with smaller heights. 
Finally, the $E_{\vec{B}}$ and $E_{\vec{B},\mathrm{free}}$ results show mixed performance: LareXd shows very good initial agreement with the GT result up to $t \approx 350$ for $E_{\vec{B}}$ and $t \approx 320$ for $E_{\vec{B},\mathrm{free}}$, but poor agreement at late time, while E1 shows poor agreement at early time but tends toward the final time GT energy values.

\subsection{Ad-Hoc Energy and Helicity Injection in Athena++ \textbf{E}-driving}\label{sec:adhoc_energy_athena}

The ``Athena ad-hoc'' series in 
Figure~\ref{fig:driving_comparison_2020_paper_Em_Emf_hr_phi}
is presented as (1) an example of how energy and helicity injection can be controlled and (2) a caution to interrogate ad-hoc and arbitrary changes to data-driving strategies. 
The simulation was terminated at $t=360$ non-dimensional units (due to a lack of computational resources) at which time the morphology could be compared to results from \citet{toriumi2020comparative}. 
These results self-consistently evolve the magnetic field, the $\mathbf{B}$ components are all coupled and do not introduce $\nabla\cdot\mathbf{B}$ errors. 
However, to better reproduce the GT magnetic energy results, the energy injection is increased by applying a greater $\nabla\cdot\mathbf{E}$. 
In this case, the correction electric field is solved using Equations~(\ref{eq:e_corrected_assimilation_curl}) and~(\ref{eq:e_corrected_driving_divE}
), the latter changed to 10\% of the divergence of the uncorrected driving electric field 
\begin{align}
    \nabla_h \cdot \boldsymbol{\mathcal{E}}^{C} = 0.1 \nabla_h\cdot \boldsymbol{\mathcal{E}}^{\textrm{vB}}.     
\end{align}
This approach can be used to account for sub-grid physics (e.g.\ photospheric motions not captured by the grid size) but there is no such justification here. 
Instead, the improved reproduction of $H_R$ and magnetic energy in the Athena ad-hoc series may indicate that there is a missing energy contribution in the E1 strategy.

\section{Conclusions}\label{sec:conclusion}
The incorporation of observed data into simulations of the solar atmosphere is becoming increasingly popular, and is described as ``data-driven modelling'' in the literature. 
In other fields the standard practice for incorporating data is \emph{data assimilation} \citep{ide1997unified}. 
Data driving uses the observations available at time $t_i$ to directly evolve the system through the boundary conditions, while data assimilation additionally tracks the forecast-observation deviation and uses it to correct the evolving state (Section~\ref{sec:driving_vs_assimilation}). 
Casting solar data driving in this standard terminology, as done in Table~\ref{table:driving_methods_literature}, is a prerequisite for importing the useful and mature assimilation and ensemble model techniques from those fields.

In the two test cases considered here, implemented in Athena++ (an Eulerian CT MHD solver), $\mathbf{E}$-driving is found to be more accurate than conventional $\mathbf{B}$-driving for reproducing flux emergence. 
The ``physically intuitive'' conventional $\mathbf{B}$-driving strategies commonly used in the literature (e.g., copying magnetograms into ghost-cells) are numerically underdetermined for an Eulerian CT scheme; imposing the desired boundary field exactly would require inverting the native update algorithm for the ghost-cell states (Section~\ref{sec:method_b_driving}),  a substantially more complex procedure than prescribing EMFs directly.
This study indicates $\mathbf{E}$-driving may be numerically advantageous, however, it remains to be seen whether the uncertainty in electric field inversions from observational data compromises this advantage when compared to the $\mathbf{B}$-driving. This is, however, difficult to definitively test since the non-inductive component of electric field cannot be constrained by observations.
For Athena++ and other Eulerian CT codes, the numerical quantity that actually advances $\mathbf{B}$ is the EMF, therefore we should specify internal EMFs through $\mathbf{E}$-driving if we intend to reproduce a desired $\mathbf{B}$ state.

The inclusion of the correction electric field (Equations~(\ref{eq:e_corrected_assimilation_curl}) and~(\ref{eq:e_corrected_driving_divE}
)) means that the $E$-driving strategy here is effectively performing data assimilation.
This illustrates the practical advantage of using the standardised terminology: recognising the correction field as an analysis step connects it to a literature in which more sophisticated treatments of the residual---e.g. Kalman-type weighting by observation and model error covariances, ensemble estimates of forecast uncertainty---are mature and could be adopted directly. This approach is already used in the ADAPT surface flux transport model~\citep{Hickmann:2015} and should be seriously considered for 3D data-driven modeling of solar coronal magnetic field evolution.


The arguments from the paragraph above can be generalised to other simulation tools---the choice of data driving strategy depends on the simulation tool's update algorithm, not only on what data are available observationally.
$\mathbf{B}$-driving approaches vary widely from (1) simply copying $\mathbf{B}$ values into ghost-cells, to (2) update-algorithm consistent inversion, but most methods described in the literature lie closer to the former. 
In principle, $\mathbf{B}$-driving could produce very accurate performance if one implemented a boundary condition that effectively inverts or otherwise exactly controls the deterministic update as is done with $\mathbf{E}$-driving in Athena++ and other codes \citep{cheung2012method,chen2026data}.
However, the development of such an inversion method poses significant challenges.
A promising alternative approach to data driving is the characteristic data-driving boundary condition by \citet{Tarr2024MHDIII}. 
However, this approach evolves a distinct MHD state rather than the target observational/GT state and so differs from our specific objective. 
A cross-code comparison with this technique and conventional data-driving strategies would be beneficial to the literature.

Last, we note that the boundary conditions on the thermodynamic variables at the driven boundary remain a limiting factor. 
The deviation in magnetic energy and magnetic free energy profiles for the GT simulations indicate under-injection of magnetic energy. 
This occurs despite an otherwise accurate reproduction of the magnetic field. 
The $\mathbf{E}$-driving cases (E1, E2, and E3) use identical density and pressure boundary conditions but vary in their velocity boundary condition. 
Considering they each use the same driving EMFs, the variation in their magnetic energy/free-energy metrics implies an inadequate thermodynamic flux treatment. 
Additionally, the density and pressure boundary conditions are often simplistic (e.g., constant hydrostatic equilibrium ghost-cell profiles that are invariant across the driven boundary).
In general, it is important to improve the thermodynamic data-driving boundary conditions because they contribute to relevant physical processes (e.g. flux-tube buoyancy) and appear to influence the energy injection in this work.  

\begin{acknowledgments}
     The authors gratefully acknowledge the support for this research through an Australian Research Council Discovery Project DP230101240.
     
	We thank Associate Professor Shinsuke Takasao for sharing his dual energy formalism implementation and Boris correction implementation in Athena++ and for providing additional details for comparison to his ground-truth simulations \citep{toriumi2020comparative}.

    We thank Associate Professor Shin Toriumi for providing guidance on his comparison paper \citep{toriumi2020comparative} and providing data from the GT simulation and figures.
	
	We acknowledge the support of computing resources and infrastructure from NCI Gadi, OzSTAR Supercomputing (Swinburne University), Sydney Informatics Hub, and the ASTAC scheme.
	Additionally, we acknowledge the advice from Professor Yang Guo, Dr Yuhong Fan, Dr Sarah Gibson, Dr Feng Chen, and Isabel Hardy. 
\end{acknowledgments}

\begin{figure}
    \includegraphics[width=\linewidth]{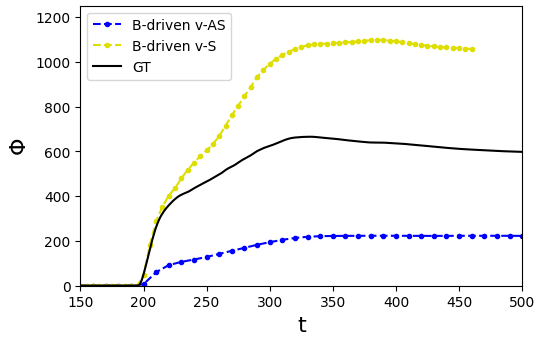}
	\caption{Total unsigned flux on the driven boundary for $\mathbf{B}$-driving with zero mass-flux (v-AS) and zero-gradient (v-S) on driving velocity. This poor reconstruction of the ground-truth (GT) results shows the important of physically meaningful velocity boundary conditions for B-driving.  \label{fig:B_driving_velocity_comparison_phi}}
\end{figure}

\begin{contribution}
K. Tapinou completed the Athena++ code development and simulations, and was the primary investigator/author of the manuscript.
A. Raboonik completed the LareXd data driving implementation and simulation, and contributed to analysis, debugging of the code, and manuscript preparation/editing. 
D. Pontin advised throughout the work, and contributed to simulation design, analysis, and manuscript preparation/editing. 
M. Wheatland advised throughout the work, and contributed to simulation design, analysis, and manuscript preparation/editing. 
M. Cheung advised throughout the work, and contributed to simulation design, analysis, and manuscript preparation/editing. 
H. Schunker advised throughout the work, and contributed to manuscript preparation/editing.


\end{contribution}

%
%

\appendix
\vspace{-0.75cm}
\section{The Athena++ Numerical Solver}\label{sec:method_athena}
Athena++ is a popular simulation tool in the wider astrophysics community but has not been adopted in solar physics. 
In the present use case, simulation of solar flux emergence, the overlapping requirements that are provided by Athena++ are below.
Nevertheless, a number of native properties make it a beneficial tool for modelling solar flux emergence: a  constrained-transport (CT) algorithm to ensure a divergence-free magnetic field; an adaptive mesh for multi-scale modelling; a range of physics modules including but not limited to radiation, viscosity, resistivity, generalised equations of state; excellent computational scalability; a community-based, open-source model.
The governing equations for Athena++ are:
\begin{eqnarray}
	\frac{\partial \rho}{\partial t } + \nabla \cdot \left(\rho \boldsymbol{v}\right) &=& 0 \\
	\frac{\partial \rho \boldsymbol{v}}{\partial t } + \nabla \cdot \left(\rho \boldsymbol{v}\boldsymbol{v} - \mathbf{B}\mathbf{B} + P^* + \Pi \right) &=& 0 \\
	\frac{\partial E}{\partial t } + \nabla \cdot \left[\left( E + P^*\right)\boldsymbol{v} - \mathbf{B}\left(\mathbf{B}\cdot \boldsymbol{v}\right)  
    \Pi\cdot \boldsymbol{v} + \eta\boldsymbol{J}\times\mathbf{B} + \boldsymbol{Q}\right] &=& 	0\\
    \frac{\partial \mathbf{B}}{\partial t}
- \nabla \times \left[
(\mathbf{v} \times \mathbf{B})
- \eta \mathbf{J}
- \frac{\eta_{\mathrm{AD}}}{|\mathbf{B}|^2}\,\mathbf{B} \times (\mathbf{J} \times \mathbf{B})
\right]
&=& 0
\end{eqnarray}
where $\rho$ is the mass density, $\boldsymbol{v}$ is the plasma velocity, $\mathbf{B}$ is the magnetic field, $P^* = P + |\mathbf{B}|^2/2$ is the total pressure, $\Pi$ is the viscous stress tensor, $E$ is the total energy density, $\boldsymbol{J}=\nabla\times\mathbf{B}$ is the current density, $\eta$ and $\eta_{\mathrm{AD}}$ are the Ohmic and ambipolar diffusivities, respectively, and $\boldsymbol{Q}$ is the heat flux.
The total energy is defined as $E = e + \rho \frac{v^2}{2} + \frac{B^2}{2}$
where $e$ is internal energy;
Pressure is calculated as $p = \left(\gamma-1\right)\left(E -\rho \frac{v^2}{2} - \frac{B^2}{2}\right)$. 
$P^\ast$ is a diagonal tensor with components $P^\ast = P + B^2/2$ (with $P$ being the gas pressure). 
These equations are written in the Heaviside-Lorentz units such that the magnetic permeability $\mu = 1$.
The interested reader is directed to the Athena++ methods paper \citep{stone2008athena,stone2020athena++} for further details. 
Note in the present work, viscosity and thermal conduction are note modelled/used.

The numerical fluxes on the lower $z$-boundary, which are vital for understanding data driving, are given by the flux vector $\boldsymbol{H}$ in Equation~(\ref{eq:H_flux_vector}). 
Fluxes in the $x$- and $y$-directions are $\bf F$ and $\bf G$, respectively.



Separate to changes required for the data driving implementation (Section~\ref{sec:data driving}) we made several generic augmentations to Athena++ to enable modelling of solar flux emergence and eruptions. 
To handle small $\beta$ solar plasmas the Boris correction \citep{gombosi2002semirelativistic} was required to limit unphysically large Alfv\'{e}n speeds that would otherwise restrict time steps to computationally intractable values. 
The Boris correction was implemented by  \cite{takasao2025connecting} in Athena++ using the formulation of \cite{matsumoto2019new}.

The well-known \citep{balsara1999maintaining,perthame1996positivity} pressure positivity issue associated with total energy conserving schemes (as is the case in Athena++) was mitigated using the dual-energy formalism \citep{bryan1995piecewise} and implemented in Athena++ by \citet{takasao2022three}. 
Conservation of total energy 
is especially important when simulating solar eruptions, where there are significant exchanges between magnetic, kinetic, and internal energies.

The driving data are loaded into Athena++ at runtime using existing the HDF5 \textbf{c++} library. 
Internally, Athena++ possesses the infrastructure to store user-data on each CPU. 
In the current implementation, the driving-data is read into memory and the relevant subset of data is stored on each CPU where appropriate. 
The relevant subset of the driving data includes ghost-cells and holds the maximum resolution possible for the driving data. 
Only mesh subsets, referred to henceforward  as meshblocks, adjacent to the lower $z$-boundary require driving data (the driven boundary). 

In the case of adaptive mesh-refinement (AMR) the driving data (already interpolated from the raw ground-truth simulation or observations) must be stored and used for a second bi-linear interpolation to the adaptive meshes (either refining or coarsening). 
This procedure occurs with the creation of each new meshblock and is internal to Athena++. 

\section{Non-dimensionalisation in Athena++}\label{sec:nondims}
The non-dimensional units used in Athena++ follow typical approaches for the hydrodynamic variables but differ for the electromagnetic variables. 
Here ``hydrodynamic properties'' refers to all fluid properties including energy density and other associated quantities. 
To characterise the hydrodynamic quantities, three independent scales are set (here we use a hat symbol to denote non-dimensional quantities and subscript zero to denote reference quantities). 
\begin{eqnarray}
	\mathbf{x} = \hat{\mathbf{x}} \, x_0, \quad  t =	\hat{t} \, t_0,  \quad 
	\rho =\hat{\rho} \, \rho_0,   \quad  \mathbf{v} = \hat{\mathbf{v}} \, x_0 \, / \, t_0, \quad 	p = \hat{p} \,  \rho_0  \,  \frac{x_0^2}{t_0^2},\;  \quad e = \hat{e} \, \rho_0 \, \frac{x_0^2}{t_0^2},
	\textrm{and}&\eta = \hat{\eta}  \, \frac{x_0^2}{t_0}.
\end{eqnarray} 
where $x_0$, $t_0$ , and $\rho_0$ are the independent scales used that represent the reference length, time, and density, respectively. 
The dimensional quantities are length ($\mathbf{x}$), density ($\rho$), velocityt ($\mathbf{v}$), pressure ($p$), internal energy ($e$), and electrical resistivity ($eta$).
Mass and temperature are treated differently since the temperature is related to thermal velocity by the particle mass treated as 
\begin{eqnarray}
	&m = \hat{m} \,  \mu_m m_p \;\textrm{and}\; &T = \hat{T} \, \frac{p_0}{k_B \, \rho_0 \, / (\mu_m m_p) }. 
\end{eqnarray} 
Note the non-dimensionalised units in Athena++ assume $\frac{k_B}{\mu_m m_p} = 1$. 

For the electromagnetic variables, Athena++ is based on Heaviside-Lorentz units which are a rationalised unit system and so no terms of $4\pi$ appear in the evolved equations, e.g., the Alfv\'{e}n speed is given by $v_A = \frac{|\mathbf{B}|}{\sqrt{\rho}}$. 

Converting from CGS Gauss ($\mathbf{B}_{\text{G}}$) to Heaviside-Lorentz Gauss ($\mathbf{B}_{\text{HLG}}$) units is done using $\mathbf{B}_{\text{G}} = \sqrt{4\pi}\mathbf{B}_{\text{HLG}}$. 
If choosing to derive the reference magnetic field from specified thermodynamic reference properties then $\mathbf{B}_{\textrm{HLG}}=\sqrt{\rho_0} x_0/t_0$  and $\mathbf{B}$ is non-dimensionalised using $\mathbf{B} = \hat{\mathbf{B}} \mathbf{B}_{\textrm{HLG}} = \hat{\mathbf{B}} \sqrt{\rho_0} x_0/t_0.$ 









\section{$\mathbf{B}$-driving Velocity Conditions}\label{sec:Bdriving_innapropriate}
To demonstrate the sensitivity of ghost-cell $\mathbf{B}$-driving strategies, we show results with zero mass-flux and zero-gradient boundary conditions on the velocity.
The driving data for magnetic field is applied in the same way as case $\mathbf{B1}$ (constant in the ghost-zones with respect to height). 
Comparing 
Figure~\ref{fig:B_driving_velocity_comparison_phi}
with Figure~\ref{fig:driving_comparison_Em_Emf_Hr_Phi} ($\Phi$)
shows $\mathbf{B}$-driving is much more sensitive to the velocity boundary condition than $\mathbf{E}$-driving. 
As explain in Sections~\ref{sec:method_b_driving} and~~\ref{sec:method_e_driving}, the magnetic field evolution is dependent on the velocity as well as the magnetic field in the boundary. 
Therefore, arbitrary conditions on the velocity boundary condition are non-physical and numerically ineffective for $\mathbf{B}$-driving, this is why they are not included in the comparison within the body of this work.   



\section{Violations of Numerical Methods and MHD Consistency by Data Driving}\label{sec:data_drivinmg_numerical_mhd_errors}

The numerical and physical accuracy of data driving can be compromised when the driving strategy only specifies a subset of the numerical fluxes on the driven boundary (Section~\ref{sec:frame_data_driving}), e.g. only specifying numerical fluxes associated with the magnetic field to achieve a desired physical magnetic-flux emergence from magnetogram observations. 
In practice, this compromise can be caused by imposing boundary conditions through direct specification of a subset of numerical fluxes (e.g.\ EMFs) or by overwriting domain values of the magnetic field. 
In both cases, the magnetic field evolution on the boundary is inconsistent with the evolution of the thermodynamic properties as anticipated by the flux vector $\mathbf{H}$ (density, momentum, and energy fluxes in Equation~(\ref{eq:H_flux_vector})) and risks generating spurious features (currents, Lorentz forces, and Poynting flux imbalances) and violating conservation properties of the numerical scheme (implicit non-conservative source terms are generated in subsequent updates within the domain). 
Data driving can avoid these issues by prescribing ghost-zone states only or with comprehensive observational data specifying all numerical fluxes, but there is still the possibility of MHD inconsistency if the ghost-zone state is not appropriately evolved. 
MHD consistency can only be guaranteed if the evolution of the driving state is dynamically realisable, i.e.\ it respects the decomposition into characteristic modes and can be representable as combinations of MHD eigenmodes \citep{tarr2024simulating}. 
At the time of writing, only \citet{Tarr2024MHDIII} uses a characteristic method to enforce MHD consistency of the driving state but this approach evolves a magnetic field state distinctly different from the target state. 
Implementation of data driving in solar physics is an exercise in compromise. 

The Athena++ $\mathbf{B}$-driving implementation is a conservative data driving scheme. 
If the ghost-zone state used to drive the simulation is not MHD consistent (in regard to the time evolution), which is the case in this work, then errors can arise in the physical realisation, but the conservation properties of the scheme are maintained. 
The $\mathbf{E}$-driving implementation is not guaranteed to produce an MHD consistent evolution or maintain strict conservation of the numerical scheme. 
The EMFs specified for the $\mathbf{E}$-driving are only calculated to produce a desired magnetic field evolution and represent specification of only a subset of the numerical fluxes. 
Furthermore, the effective driving state has not been explicitly calculated to form an MHD consistent state. 
For an extended discussion on this topic, see \citet{tarr2024simulating}.

The LareXd $\mathbf{B}$-driving implemented here is susceptible to similar problems as the Athena++ $\mathbf{E}$-driving implementation and an additional magnetic field inconsistency. 
The MHD evolution in the Lagrangian step is inherently conservative without the use of numerical fluxes. 
However, the Lagrangian remapping procedure to the Eulerian grid must be completed while explicitly maintaining conservation (using fluxes). 
Enforcing $B_z$ and $\boldsymbol{v}$ on the lower $z$-boundary (which is part of the simulation domain) therefore risks producing an implicit non-conservative source term and an MHD-inconsistent state. 
Enforcing domain values bypasses the conservative remap and disrupts the coupling between the Lagrangian and remapped states. 
Additionally, $\nabla\cdot\mathbf{B}$ errors are introduced at the lower $z$-boundary because enforcing $B_z$ decouples it from the self-consistently evolved $B_x$ and $B_y$ values.

The objective in this work is to achieve a desired magnetic-flux emergence and magnetic field configuration from incomplete driving data. 
This objective is accomplished in our implemented driving strategies, though we highlight the possible numerical and physical accuracy errors that may arise in the implementations presented. 
We suspect that these errors are reduced by the high cadence of the driving data and by the proximity of the driving data to an MHD-consistent state. 
In contrast, low-cadence data may impose large changes between consecutive
driving frames and nonphysical local gradients near the boundary, producing impulsive boundary forcing and strong transients.
This discussion and the results comparing driving strategies demonstrate the compromises that must be made in the absence of spatially complete driving data and strictly correct (but challenging) boundary treatments \citep[e.g.][]{Tarr2024MHDIII}. 

\vspace{-0.75cm}

\bibliography{publication_references_AJL}{}
\bibliographystyle{aasjournalv7}



\end{document}